\documentclass[aip,
 amsmath,amssymb,
 reprint,%
]{revtex4-1}

\newcommand{\sech}{\mathrm{sech} \,}
\usepackage{graphicx}% Include figure files
\usepackage{dcolumn}% Align table columns on decimal point
\usepackage{bm}% bold math
\usepackage{textcomp}
\usepackage{amsmath}
\usepackage{verbatim}
\usepackage{comment}
\usepackage{lipsum}
\usepackage{xcolor}
\usepackage{soul}
\usepackage{multirow}
\usepackage{makecell}
\usepackage{rotating}
\usepackage{subcaption}
\usepackage{ragged2e}
\begin{document}

\title[]{Dynamics and modulation of cosmic ray modified magnetosonic waves in a galactic gaseous rotating plasma}% Force line breaks with \\
\author{Jyoti Turi}
\email{ jyotituri.maths@gmail.com}
\affiliation{Department of Mathematics, Siksha Bhavana, Visva-Bharati University, Santiniketan-731 235, West Bengal, India}
\author{Gadadhar Banerjee}
\email{gadadhar@burdwanrajcollege.ac.in, gban.iitkgp@gmail.com}
\affiliation{Department of Mathematics, Burdwan Raj College, University of Burdwan, Burdwan-713 104, West Bengal, India}
\date{\today}% It is always \today, today,

\begin{abstract}
The influence of the presence of cosmic fluid on the magnetosonic waves and modulation instabilities in the interstellar medium of spiral galaxies is investigated. The fluid model is developed by modifying the pressure equation in such dissipative rotating magnetoplasmas incorporating thermal ionized gas and cosmic rays. Applying the normal mode analysis, a modified dispersion relation is derived to study linear magnetosonic wave modes and their instabilities. The cosmic rays influence the wave damping by accelerating the damping rate. The standard reductive perturbation method is employed in the fluid model leading to a Korteweg–de Vries-Burgers (KdVB)  equation in the small-amplitude limit. Several nonlinear wave shapes are assessed by solving the KdVB equation, analytically and numerically. The cosmic ray diffusivity and magnetic resistivity are responsible for the generation of shock waves. The modulational instability (MI) and the rogue wave solutions of the magnetosonic waves are studied by deriving a nonlinear Schr\"{o}dinger (NLS) equation from the obtained KdVB equation under the assumption that the cosmic ray diffusion and magnetic resistivity are weak and the carrier wave frequency is considerably lower than the wave frequency. The influence of various plasma parameters on the growth rate of MI is examined. The modification of the pressure term due to cosmic fluid reduces the MI growth in the interstellar medium. In addition, a quantitative analysis of the characteristics of rogue wave solutions is presented. Our investigation's applicability to the interstellar medium of spiral galaxies is traced out.
\end{abstract}

\maketitle

%%%%%%%%%%%%%%%%%%%%%%%%%%%%%%%%%%%%%%%%%%%%%%%%%%%%%%
\section{INTRODUCTION} \label{Sec-Intro}
In our Galaxy, interstellar medium (ISM) clouds are highly inhomogeneous and comprise various phases of gas clouds, such as a hot phase, a warm phase, diffusive clouds and molecular clouds. These components of the ISM cloud exhibit a similar energy density distribution\cite{kuwabara2020dynamics}. It is well-known that the galactic cosmic rays significantly contribute to the evolution and structure formation of ISM clouds. Galactic cosmic rays \cite{parker1969galactic,padovani2020impact} are highly energetic particles ejected from astrophysical bodies outside our solar system, such as supernova remnants, planetary disks and the Sun, which numerously interact with different components of the ISM. Cosmic rays play an important role in the charge of dust grains \cite{kalvans2016temperature}, gas ionization \cite{yusef2007cosmic}, and energy transfer \cite{gaches2019astrochemical} in the interstellar medium. They interact with cosmic dust grains via different physical mechanisms, leading to the generation of different wave modes and instabilities in the media. They are major contributors to the ionization of molecular clouds (MCs) core \cite{padovani2011effects} and the ionization of molecule hydrogen ($H_2$)  in diffuse interstellar clouds \cite{indriolo2009implications}. They transfer energy to the dust grains through heating the dust grains in cosmic ray-plasma interactions.

The nonlinearities in the plasma contribute to energy localization, leading to the evaluation of various types of nonlinear coherent structures, viz., solitons and shocks, which have immense application in space as well as astrophysical plasma environments\cite{goertz1989dusty,havnes1992charged, horanyi1986effects}. The evolutions of these nonlinear coherent structures are characterized by various well-known nonlinear partial differential equations, such as Korteweg-de Vries (KdV), Korteweg–de Vries–Burgers (KdVB), Zakharov-Kuznetsov (ZK), Kadomtsev-Petviashvili (KP) equations and their families\cite{wazwaz2017abundant,chen2018special,el2016generation}. Another crucial equation that controls the motion of nonlinear structures is the nonlinear Schrödinger (NLS) equation\cite{mihalache2017multidimensional}. A rogue wave\cite{akhmediev2009waves,muller2005rogue} is one of the viable rational solutions of the NLS equation\cite{sharma2024modulation,chen2015vector}. Such waves are often referred to as monster, violent, extreme, or giant waves that have a few times higher amplitude than solitary waves and are observed in coastal waters \cite{kharif2008rogue}, fibre optic and optical medium \cite{solli2007optical}, Bose-Einstein condensates \cite{bludov2009matter}, atmosphere \cite{stenflo2010rogue} and astrophysical environments \cite{sabry2012freak}. There has been increasing interest in investigating the behaviour of magnetosonic waves and the associated coherent structures, viz., solitons, shocks and rogue waves in different plasma environments\cite{rahim2019magnetosonic,hager2023magnetosonic,el2012rogue,moslem2011surface,akbari2014electrostatic}.

In recent times, researchers have devoted significant attention to the role of cosmic rays in various wave modes in linear regimes and their stability/instabilities. To mention a few, in a recent work, Marcowith \textit{et al.} \cite{marcowith2021cosmic} have addressed Cosmic ray-driven streaming instability in different space and astrophysical plasmas. Zhou \textit{et al.} \cite{zhou2024effects} have studied the impact of cosmic rays on the dynamics of solar wind in the outer heliosphere, considering three components of magnetohydrodynamic (MHD) plasma consisting of the solar wind, interstellar neutral atoms, and cosmic rays. The study shows that cosmic rays may decrease the speed of the solar wind shocks and affect the shock structure. In this direction, Turi and Misra \cite{turi2022magnetohydrodynamic} have addressed the impact of cosmic rays on the propagation of MHD waves as well as their instability (Jeans) for a self-gravitating galactic gaseous cloud. Mansuri \textit{et al.} \cite{mansuri2023evolution} have theoretically studied the Jeans instability considering cosmic rays in a radiative quantum plasma, including Hall current and Coriolis force. It is reported that cosmic rays significantly alter the excitation of Jeans instability and affect the instability region as well as the instability growth rate. In another work, Boro \textit{et al.} \cite{boro2023cosmic} have studied the Jeans instability for the molecular dusty plasma environment of molecular clouds, including cosmic rays. Recently, some new exciting features of low-frequency MHD waves and associated instability (Jeans) consisting of superthermal gas and cosmic rays in an astrophysical plasma environment have been addressed by Boro and Prajapati \cite{boro2024suprathermal}. Further, various previous studies suggest that the Coriolis force plays an important role in various plasma phenomena in different plasma environments, such as rotating plasma in the laboratory as well as astrophysical plasma\cite{turi2022magnetohydrodynamic, hager2023magnetosonic}. 

Existing literature reveals that there is still some lack of studies in the direction of theoretical investigation of magnetosonic waves in the ISM of spiral galaxies, exploring the effect of cosmic rays' interaction with ionized gas in such dissipative rotating magnetoplasma. The main objective of the present study is to advance the theory of magnetosonic wave dynamics in the ISM of spiral galaxies, consisting of thermally ionized gas and cosmic rays. The cosmic rays are assumed to interact with the thermally ionized gas in a fluid-fluid approach and modify the existing pressure law of the plasma. We aim to explore the impact of cosmic rays, in terms of the parameters related to the cosmic ray pressure and diffusion, on both the linear and nonlinear magnetosonic waves. Toward this objective, we use the basic MHD equations including the cosmic ray modified pressure law for a fairly good approximation in studying the structures and evolution of the magnetosonic waves. The model is suitable and well justified in real physical situations of ISM clouds in spiral galaxies.

 The manuscript is organised as follows: in Sec. \ref{sec-bas-eq}, we describe the basic equations for the MHD waves. In Sec. \ref{Sec-Linear}, linear properties of the magnetosonic waves are discussed through the linear dispersion relation. Sec. \ref{sec-nls} includes nonlinear analysis, where we derive a Korteweg–de Vries–Burgers (KdVB) equation for the magnetosonic waves and analyse the analytical and numerical behaviour of the equation. Sec. \ref{Sec-MI} is devoted to modulational instability and rogue wave solutions for carrier wave frequencies much smaller than the wave frequency. A sensitivity analysis for the plasma parameters is performed in Sec. \ref{sec-sensitivity}. Finally, we summarize our main findings in Sec. \ref{sec-result}.

  %%%%%%%%%%%%%%%%%%%%%%%%%%%%%%
\begin{figure*}
\centering
 \begin{subfigure}[t]{0.32\textwidth}
     \centering
      \includegraphics[width=\textwidth,height=2 in]{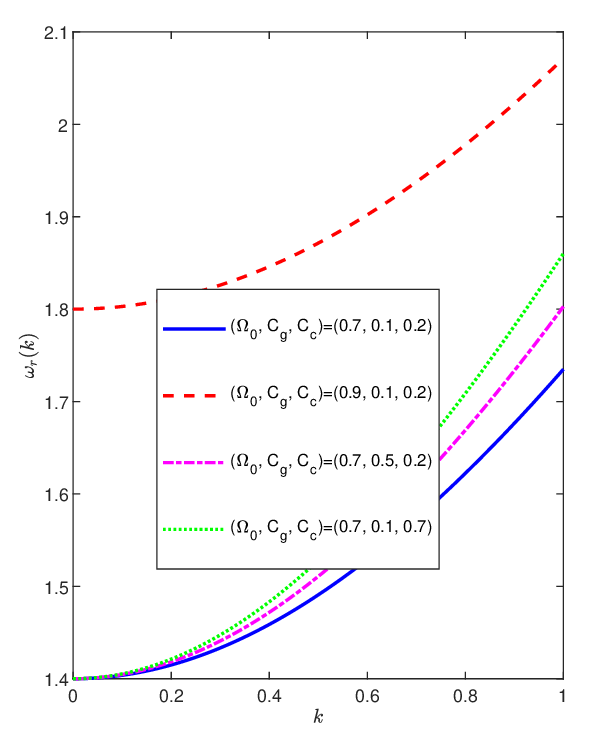}
     \caption{}
     \label{dis-a}
 \end{subfigure}
 \hfill
 \begin{subfigure}[t]{0.32\textwidth}
     \centering
     \includegraphics[width=\textwidth,height=2.1 in]{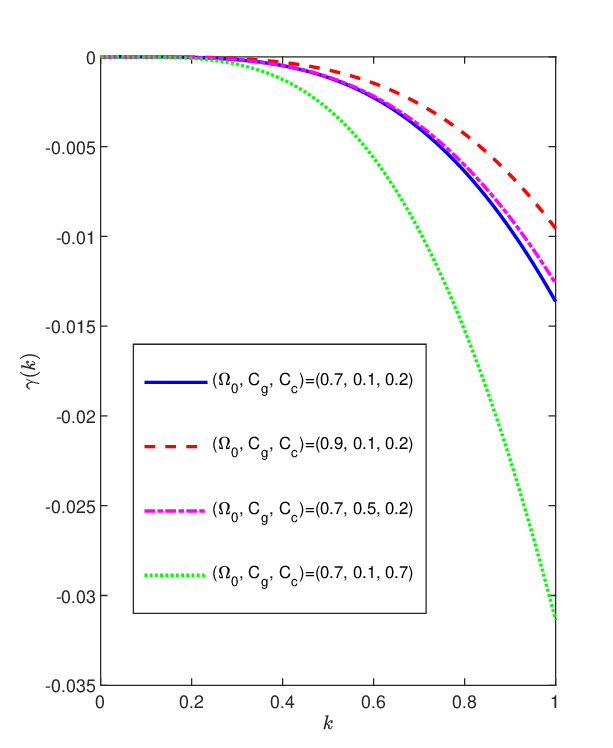}
     \caption{}
     \label{dis-b}
 \end{subfigure}
 \hfill
 \begin{subfigure}[t]{0.32\textwidth}
     \centering
      \includegraphics[width=\textwidth,height=2.1 in]{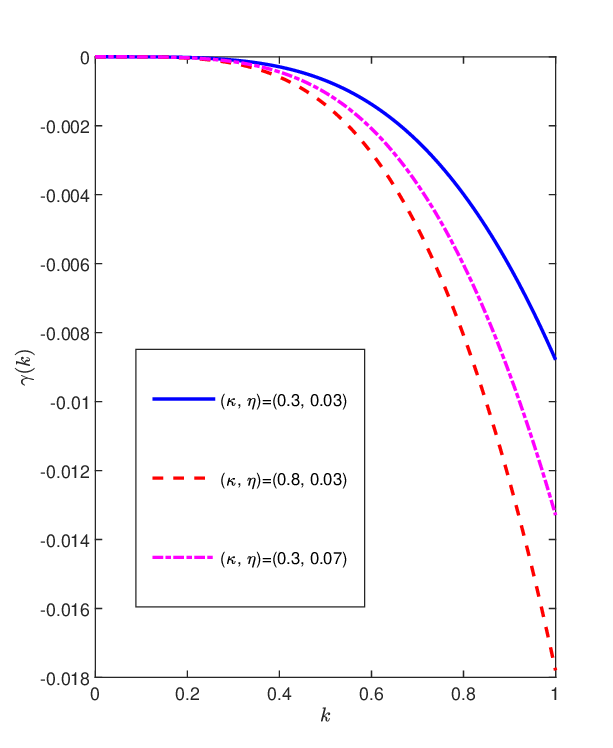}
     \caption{}
     \label{dis-c}
 \end{subfigure}
    \caption{\justifying The real part $\omega_r \equiv \omega_r(k)$ and the damping rate $\gamma \equiv \gamma(k)$ of the dispersion relation \eqref{eq-dis-parallal}  are displayed for different values of the parameters as mentioned in the legends. The profiles show that, for $k \ll 1$, the wave frequency ($\omega_r$) remains constant and the wave damping ($\gamma$) is negligible. Beyond the limit, both increase for increasing values of $k$. The damping occurs due to the effects of the parameters $C_g$ and $C_c$ related to thermal and cosmic ray pressure, rotational frequency ($\Omega_0$), cosmic ray diffusivity ($\kappa$) and magnetic resistivity ($\eta$). Here, ($\theta$, $\kappa$, $\eta$)= ($\pi/3, 0.3, 0.03)$ in the subplots (a) \& (b)  and $(\theta, \Omega_0, C_g, C_c) = (\pi/3, 0.9, 0.2, 0.4)$ in the subplot (c). These dimensionless typical plasma parameters for the ISM of spiral galaxies are chosen from Table I.}
    \label{fig-dispersion}
\end{figure*}

%%%%%%%%%%%%%%%%%%%%%%%%%%%%%%%%%%%%%%

\section{BASIC EQUATIONS OF THE PROBLEM} \label{sec-bas-eq}
The typical plasma environment of the interstellar medium, composed of ionized thermal gas and cosmic rays, has been considered to study the propagation of magnetosonic waves. Here, the thermal ionized gas is considered polytropic, having no diffusion in any direction, and cosmic rays are assumed to be gas with negligible density but contributing significantly to modify the pressure of the plasma environment \cite{bonanno2008generation,turi2022magnetohydrodynamic}. The plasma is supposed to be immersed in a uniform external magnetic field $\mathbf{B}=B(x,t) \hat{e}_z$, where $B(x,t)$ is the static external magnetic field strength and $\hat{e}_z$ denotes the unit vector along the $z$ direction. The interstellar medium's environment is considered to be homogeneous, fully ionized and conducting fluid. The plasma is assumed to rotate slowly with a rotational frequency $\Omega$ around an axis lying in the $xz$-plane. The basic   evolution equations describing the dynamics of magnetosonic waves  are given as \cite{turi2022magnetohydrodynamic,Turi_2024}
\begin{equation}
	\label{continuity_eq}
	\frac{\partial \rho}{\partial t} +\mathbf{\nabla}\cdot\left(\rho \mathbf{v}\right)=0,
\end{equation} 
\begin{multline}
	\label{momentum_eq}
	\frac{\partial \mathbf{v}}{\partial t} +\left(\mathbf{v}\cdot\mathbf{\nabla}\right) \mathbf{v}=-\frac{1}{\rho}\mathbf{\nabla}\left(P+\frac{\mathbf{B}^2}{2\mu_0}\right)
	\\+\frac{1}{\rho \mu_0} \left(\mathbf{B}\cdot\mathbf{\nabla}\right)\mathbf{B}-2\mathbf{\Omega}\times\mathbf{v},
\end{multline}

\begin{equation}
	\label{magnetic_eq}
	\frac{\partial \mathbf{B}}{\partial t} +\left(\mathbf{v}\cdot\mathbf{\nabla}\right) \mathbf{B}=\left(\mathbf{\textbf{B}}\cdot\mathbf{\nabla}\right) \mathbf{v}-\left(\mathbf{\nabla}\cdot\mathbf{v}\right) \mathbf{B} +\eta \mathbf{\nabla^2 B},
\end{equation}
where $\rho$ and $\mathbf{v}$ denote the fluid (thermal gas) mass density and velocity, respectively, $\mu_0$ is the vacuum permittivity, and $\eta$ is the coefficient of the magnetic resistivity. The total pressure is given by $P=P_g+P_c$ where $P_g$ and $P_c$ are the gas pressure and the pressure due to the cosmic rays, respectively. As the fluid is assumed to be a thermal gas which behaves adiabatically, we use the following equation of state,
\begin{equation}
    \frac{\partial P_g}{\partial t} + \mathbf{v}\cdot \mathbf{\nabla} P_g +\gamma_g \mathbf{\nabla} \cdot \mathbf{v}=0,
\end{equation}
where $\gamma_g$ is the specific heat ratio of the thermal gas. The diffusion convection equation for cosmic rays is given by  \cite{zank1990weakly}
\begin{equation}
\label{eq-cosmic-ray}
 \frac{\partial P_c}{\partial t} + \mathbf{v}\cdot \mathbf{\nabla} P_c +\gamma_c \mathbf{\nabla} \cdot \mathbf{v}-\kappa \mathbf{\nabla}^2 P_c=0, 
\end{equation}
where $\gamma_c$ is the adiabatic index corresponding to cosmic rays, assumed to be a constant, and $\kappa$ is the hydrodynamic form of the diffusion coefficient. % and is generally a function of position and wave energy. However, since we are considering weak shocks, we can take $\kappa$ to be a constant without violating any generality \cite{zank1990weakly}. 

The magnetosonic waves are assumed to propagate along the $x$-direction, magnetic fields are supposed to be along $z$-direction, and the plasma is slowly rotating around an axis, making an angle $\theta$ with $z$ the axis. Thus, by replacing $\mathbf{\nabla}$ by $(\frac{\partial}{\partial x}, 0,0)$, $\mathbf{v}$ by $(v_x, v_y, v_z)$ and $\mathbf{\Omega}$ by $(\Omega_0 \sin \theta, 0, \Omega_0 \cos\theta)$, we rewrite the basis Eqs. \eqref{continuity_eq}-\eqref{eq-cosmic-ray}  as follows
\begin{equation}
   \label{eq-n-con}
    \frac{\partial\rho}{\partial t} +\frac{\partial}{\partial t} \left(\rho v_x\right)=0, 
\end{equation}
\begin{equation}
	\rho \left(\frac{\partial v_x}{\partial t} +v_x \frac{\partial v_x}{\partial x}\right) =\frac{\partial P_g}{\partial x} -\frac{\partial P_c}{\partial x} - B \frac{\partial B}{\partial x} +2 \rho \Omega_0 \cos \theta v_y,
\end{equation}
\begin{equation}
	\frac{\partial v_y}{\partial t} +v_x \frac{\partial v_y}{\partial x}=2 (\Omega_{0} \sin \theta  v_z-\Omega_0 \cos \theta v_x),
\end{equation}
\begin{equation}
\frac{\partial v_z}{\partial t}+v_x\frac{\partial v_z}{\partial x}= -2 \Omega_{0} \sin\lambda  v_y,	
  \end{equation}
\begin{equation}
\label{eq-n-mag}
	\frac{\partial B}{\partial t} +\frac{\partial }{\partial x}\left(v_x B\right)= \eta \frac{\partial^2 B}{\partial x^2},
\end{equation}
\begin{equation}
   \frac{\partial P_g}{\partial t} + v_x \frac{\partial}{\partial x} P_g +\gamma_g P_g \frac{\partial v_x}{\partial x}=0, 
\end{equation}
\begin{equation}
\label{eq-n-cosmic}
   \frac{\partial P_c}{\partial t} + v_x \frac{\partial}{\partial x} P_c +\gamma_c P_c \frac{\partial v_x}{\partial x}-\kappa \frac{\partial^2 P_c}{\partial x^2}=0.
\end{equation}

The above set of Eqs. \eqref{eq-n-con}-\eqref{eq-n-cosmic} is made normalized by normalising the variables and parameters as follows:  
$\rho\rightarrow\rho/\rho_0$, $B\rightarrow B/B_0$, $t \rightarrow t \omega_{ci}$, $v_x\rightarrow v_x/V_A$, $v_y\rightarrow v_y/V_A$, $v_z\rightarrow v_z/V_A$, $x\rightarrow x \omega_{ci}/V_{A}$, $\Omega\rightarrow \Omega/\omega_{ci}$, $P_g \rightarrow P_g/\rho_0 V_A^2$, $P_c \rightarrow P_c/\rho_0 V_A^2$, $\eta\rightarrow \eta \omega_{ci}/V_A^2$, $\kappa \rightarrow \kappa \omega_{ci}/V_A^2$, 
where, $\rho_0$ is the equilibrium density, $B_0$ is the equilibrium magnetic field strength, and $\omega_{ci}=\frac{eB_0}{m_i}$ is the ion cyclotron frequency, $V_A=\frac{B_0}{\sqrt{\mu_0 \rho_0}}$ is the Alfv{\'e}n velocity. Furthermore, our analysis is suitable based on real magnetoplasma environments, which are desired to be relevant in spiral galaxies \cite{gliddon1966gravitational,turi2022magnetohydrodynamic}. Consequently, for numerical demonstration, we choose the typical consistent parameter values as given in Table I. In this regime, the ion gyro-frequency is typically scaled as $\omega_{ci}\sim 10^{-1}$ s$^{-1}$. The plasma is assumed to be rotated with small rotational frequency, i.e. $\Omega_0<1$  allowing us to ignore the second- and higher-order terms of $\Omega_0$. Hence, we neglect the effect of centrifugal force $\Omega_0 \times (\Omega_0 \times r)$\cite{abdikian2022drift,abdikian2021supernonlinear}. Denormalization of $\Omega_0$ implies $\Omega_0/\omega_{ci} \ll 1$. Thus, the numerical value of the rotational frequency is taken to be much lower than the ion gyro-frequency $\omega_{ci}$. 
%%%%%%%%%%

\begin{table} 
\begin{tabular}{|c |c|c|c|}
\hline
Physical quantity & Symbol & Value & Units\\
\hline
Fluid density &  $\rho_0$ & $2 \times 10^{-21}$ & Kg/m$^{3}$\\
Magnetic field & $B_0$ & $1 \times 10^{-9}$ & T \\
Thermal pressure & $P_{g0}$ & $2.95\times 10^{-13}$ & N/m$^2$\\
Cosmic ray pressure & $P_{c0}$ &$2.14\times 10^{-13}$ & N/m$^2$\\
Electron charge & $e$ & $1.6\times 10^{-19}$ & C \\
Ion mass& $m_i$ & $1.67 \times 10^{-27}$ & kg \\
Thermal speed & $C_g$ & $1.39\times 10^4$ & m s$^{-1}$ \\
Cosmic ray speed & $C_c$ & $1.19\times 10^4$ &  m s$^{-1}$ \\
\makecell{Rotational\\ frequency} & $\Omega_0$ & $0.076$ & s$^{-1}$\\ 
\makecell{Adiabetic index\\ of thermal gas} & $\gamma_g$ & 4/3& \makecell{Unit\\less} \\
\makecell{Adiabetic index\\ of cosmic rays} & $\gamma_c$ & 4/3& \makecell{Unit\\less} \\
\makecell{Permeability \\of free space} & $\mu_0$ & $1.26\times 10^{-6}$ & H/m \\
Magnetic resistivity & $\eta$ & $8.28 \times 10^7$ & m$^2$ s$^{-1}$  \\
Cosmic ray diffusivity & $\kappa$ & $3.72 \times 10^9$& m$^2$ s$^{-1}$ \\
\hline
\end{tabular}
\caption{\justifying Values of the typical plasma parameters of spiral galaxies.\cite{gliddon1966gravitational,turi2022magnetohydrodynamic} }
\end{table}
%%%%

%%%%%%%%%%%%%%%%%%%%%%%
 %%%%%%%%%%%%%%%%%%%%%%%%%%%%%%%%%%%%%%%%%%%%%%%%%%%%%%%%%%%%
\begin{figure*}
\centering
 \begin{subfigure}[t]{0.49\textwidth}
     \centering
     \includegraphics[width=\textwidth,height=2.5 in]{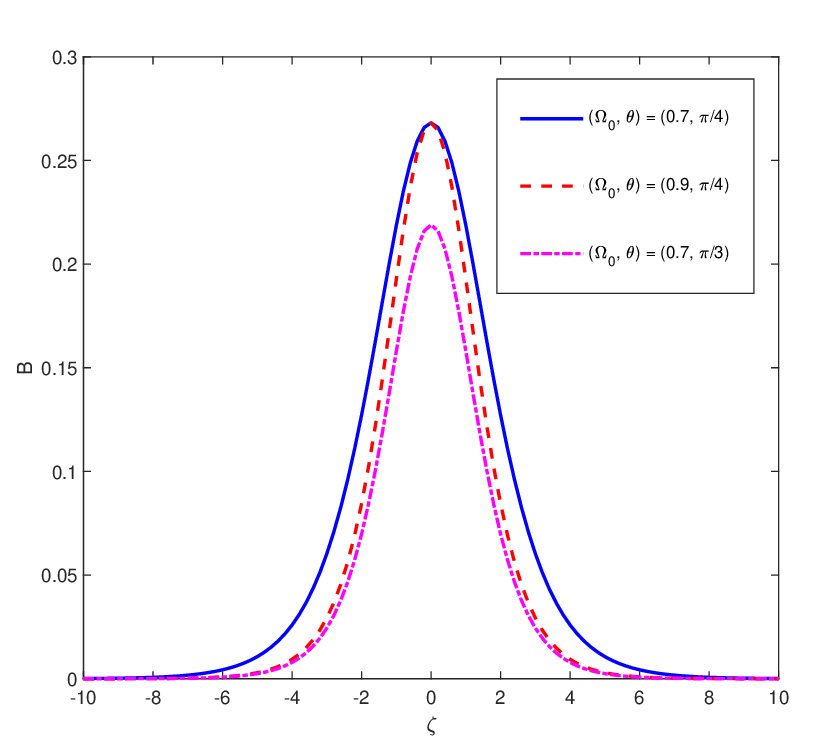}
     \caption{}
     \label{soliton-a}
 \end{subfigure}
 \hfill
 \begin{subfigure}[t]{0.49\textwidth}
     \centering
     \includegraphics[width=\textwidth,height=2.5 in]{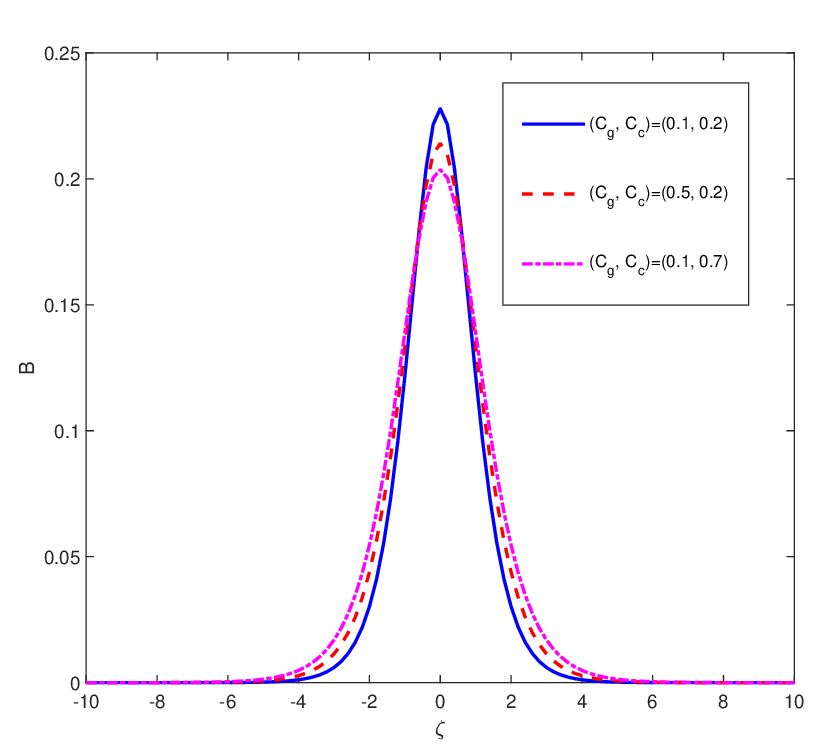}
     \caption{}
     \label{soliton-b}
 \end{subfigure}
    \caption{\justifying The typical one-soliton profiles $B ~\text{vs}~\zeta$ are plotted [see Eq. \eqref{eq-1soliton-solu}] for different values of the parameters as mentioned in the legends. The wave amplitude remains unchanged as the rotating frequency $\Omega_0$  increases, however, it decreases due to the increasing value of the angle of rotation ($\theta$) and parameters related to thermal pressure ($C_g$) and cosmic ray pressure ($C_c$). The width of the waves decreases due to the increasing effect of  $\Omega_0$ and $\theta$, but it becomes wider as $\theta$, $C_g$ and $C_c$ increase. Here, $(C_g, C_c)$= $( 0.2, 0.4)$ and $(\Omega_0, \theta) = (0.9, \pi/3)$ in the subplots (a) and (b), respectively, and $v_0=0.1$. The other parameters remain unchanged as in Fig. \ref{fig-dispersion}. }
    \label{Fig-solitary-profile}
\end{figure*}
%%%%%%%%%%%%%%%%%%%%%%%%%%%%%%%%%%%%%%%%%%%%%%%%%%%%

%%%%%
\section{LINEAR ANALYSIS} \label{Sec-Linear}
We study the propagation of magnetosonic plane waves in this plasma medium along the $x$-direction. To proceed, we linearize the set of Eqs. \eqref{eq-n-con}-\eqref{eq-n-cosmic} about the equilibrium state, assuming the plasma is uniform with constant density and zero velocity. Dividing the physical quantity $f$ into the equilibrium ($f_0$) and perturbed ($f_1$) parts as $f=f_0+f_1$, where $f=(\rho,B,v_{x},v_{y},v_{z},P_{g},P_{c})$, $f_0=(1,1,0,0,0,P_{g0},P_{c0})$ and $f_1=(\rho_1,B_1,v_{x1},v_{y1},v_{z1},P_{g1},P_{c1})$, we obtain a linearized set of equations from Eqs. (\ref{eq-n-con})-(\ref{eq-n-cosmic}).

 Assuming the perturbations to vary as a plane wave of the form $\sim exp\left(i kx-i \omega t\right)$ with wave number $k$ and wave frequency $\omega$, we get the following linear dispersion relation
\begin{eqnarray}
\label{eq-linear-dis-rel}
  \omega^2=\left(C_{g}^2+C_{c}^2\frac{\omega}{\omega+i \kappa k^2} + \frac{\omega}{\omega+i \eta k^2}\right) k^2\nonumber\\
  + \frac{4\Omega_0^2 \cos^2 \theta \omega^2}{(\omega^2- 4\Omega_0^2 \sin^2 \theta)},
\end{eqnarray}
where $C_{g} (=\sqrt{\gamma_g P_{g0}})$ and  $C_{c} (=\sqrt{\gamma_c P_{c0}})$ are the normalized form (normalized by $V_A$) of the speed of thermal gas and cosmic rays, respectively. It is worth mentioning that the denormalization of the above quantities yields $C_g=\sqrt{\frac{\gamma_g P_{g0}}{\rho_0}}$ and $C_c=\sqrt{\frac{\gamma_c P_{c0}}{\rho_0}}$. Due to the presence of the parameters $\eta$ and $\kappa$ in the dispersion relation, $\omega$ and $k$ become complex. The dispersion is also influenced by the rotational frequency ($\Omega_0$) and the angle $\theta$ related to the rotation of the fluid in the $xz$-plane.
If the axis of rotation is taken along $x$-axis (i.e. $\theta=\pi/2$), the dispersion relation \eqref{eq-linear-dis-rel}  modified to
\begin{eqnarray} \label{eq-dis-pi2}
  \omega^2=\left(C_{g}^2+C_{c}^2\frac{\omega}{\omega+i \kappa k^2} +\frac{\omega}{\omega+i \eta k^2}\right) k^2,
\end{eqnarray}
where the effects of the Coriolis force disappear due to the rotation of the plasma along the $x$-axis, however, the influences of magnetic resistivity and cosmic ray diffusivity have been retained. The wave becomes dispersionless in long-wavelength limits (i.e. when $k \ll 1$) and/or when the effects of magnetic resistivity and cosmic ray diffusivity are weaken. Consequently, the dispersion reduces to $\omega=ak$ where $a=(1+C_{g}^2+C_{c}^2)^{1/2}$. 
In the case of when the axis of rotation is taken along the $z$-axis (i.e. $\theta=0$), the dispersion relation \eqref{eq-linear-dis-rel} reduces to
\begin{eqnarray} 
\label{eq-dis-parallal}
  \omega^2=\left(C_{g}^2+C_{c}^2 \frac{\omega}{\omega+i \kappa k^2} + \frac{\omega}{\omega+i \eta k^2}\right) k^2+4\Omega_0^2.
\end{eqnarray}
The dispersion \eqref{eq-dis-parallal} includes the effect of the Coriolis force due to the rotation of the plasma. Assuming the plasma to be a highly conducting fluid, if the effect of magnetic resistivity is ignored, the above-normalized dispersion relation \eqref{eq-dis-parallal} reduces to the dispersion relation as obtained by
% \begin{eqnarray} 
% \label{eq-dis-parallal-denor}
%   \omega^2=\left(C_{g}^2+C_{c}^2 \frac{\omega}{\omega+i \kappa k^2} + V_A^2\right) k^2+4\Omega_0^2.
% \end{eqnarray},
Turi and Misra\cite{turi2022magnetohydrodynamic} upon denormalization, in the absence of self-gravitation. Further, disregarding the effect of Coriolis force due to the rotation of the fluid, cosmic ray pressure, and cosmic ray diffusion, one can recover the typical magnetosonic mode with phase velocity $\omega/k=(C_g^2+V_A^2)^{1/2}$ as given in Bittencourt \cite{bittencourt2013fundamentals}. Due to the effect of cosmic ray diffusivity and magnetic resistivity, the wave becomes unstable, i.e., it can be either damped or anti-damped. To obtain the growth rate (or damping), we separate the real and imaginary parts of the wave frequency as $\omega=\omega_r +i \gamma$, assuming $|\gamma|$, $|\kappa|$, $|\eta| \ll \omega_r$,
$\frac{\omega}{\omega+i \kappa k^2} \approx (1-i\frac{\kappa k^2}{\omega_1})$ and $\frac{\omega}{\omega+i \eta k^2} \approx (1-i\frac{\eta k^2}{\omega_1})$, where $\omega_1$ is the solution of Eq. \eqref{eq-dis-parallal} at $\eta=\kappa=0$. We found that the magnetosonic wave gets damped, and the absolute value of damping is estimated to be $|\gamma| \sim \frac{1}{2} (C_c^2 \kappa +\eta) \frac{k^4}{\omega_1^2}$ and the real part is $\omega_r=\omega_1$. The dispersion relation \eqref{eq-dis-parallal} includes complex parts due to the contribution of $\eta$ and $\kappa$ to the wave mode. In the context of spiral galaxies, a significant contribution of $\eta$ and $\kappa$ causes damping into the waves, resulting in wave energy decay, hence transferring energy to the ISM. This transformation of wave energy to thermal energy contributes to heating the ISM of spiral galaxies.

To study the dispersion and damping characteristic of the magnetosonic wave, we take numerical results from Eq. \eqref{eq-dis-parallal} displayed in Fig. \ref{fig-dispersion}.  The plots of the real part $\omega_r=\omega_r(k)$ are shown in the subplot (a) for different values of the parameters $\Omega_0$, $C_g$ and $C_c$ associated with the rotational frequency, thermal pressure and cosmic ray pressure, respectively. It is found that initially, for $k \ll 1$, the profile of $\omega_r$ remains parallel to the axis of wave number, i.e., the phase velocity approaches a constant value, as we discussed earlier. Beyond the limit $k \ll 1$, the frequency of the wave increases with increasing values of $k$. The profiles ($\omega_r$) are plotted for four sets of values of $\Omega_0$, $C_g$ and $C_c$. It has been observed from the profiles that an increase in either of these three parameters, keeping the values of alternate parameters fixed, leads to an enhancement of the wave frequency. However, in the limit $k \ll 1$, the increases in wave frequencies are found to be negligible for increasing values of $C_g$ and $C_c$, beyond the limit, the increments are transpicuous. In the subplots (b) and (c) of Fig. \ref{fig-dispersion}, the profiles of estimated damping rate $\gamma=\gamma(k)$ are plotted for different values of $\Omega_0$, $C_g$, $C_c$ and $\kappa$, $\eta$, respectively. The figures show that the absolute damping rate is negligible in the limit $k \ll 1$, but the damping grows with increasing values of $k$. Subplot (b) depicts that an increase in the values of $\Omega_0$ and $C_g$ decreases the absolute damping rate, whereas the wave damping increases with increasing value of the parameter $C_c$. On the other hand, from the subplot (c), the significant growth in the damping rate has been noticed with the increasing values of $\eta$ and $\kappa$, which leads to unstable behaviour on wave mode. Magnetic resistivity $\eta$ and cosmic ray diffusivity $\kappa$ do not contribute to the real part of the wave frequency but influence the wave damping. However, subplots (a) and (b) show that the absolute damping $\gamma$ reduces as the rotating frequency increases, but the real part of the wave frequency increases. So, it is interesting to observe that the damping due to magnetic resistivity and cosmic ray diffusivity becomes insignificant with an increase in rotating frequency. 
%%%%%%%%%%%%%%%%%%%%%%%%%%%%%%%%%%%
\begin{figure*}
\centering
 \begin{subfigure}[t]{0.19\textwidth}
     \centering
     \includegraphics[width=\textwidth,height=2.05 in]{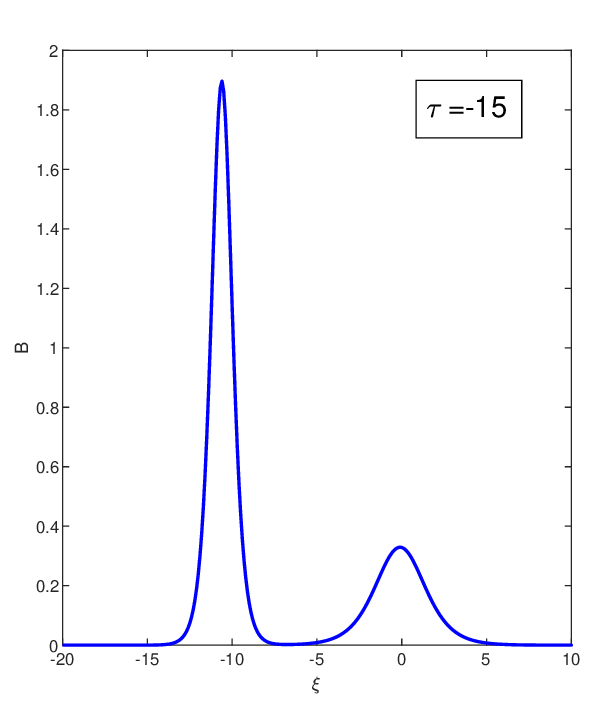}
     \caption{}
     \label{2soliton-a}
 \end{subfigure}
 \hfill
 \begin{subfigure}[t]{0.19\textwidth}
     \centering
     \includegraphics[width=\textwidth,height=2 in]{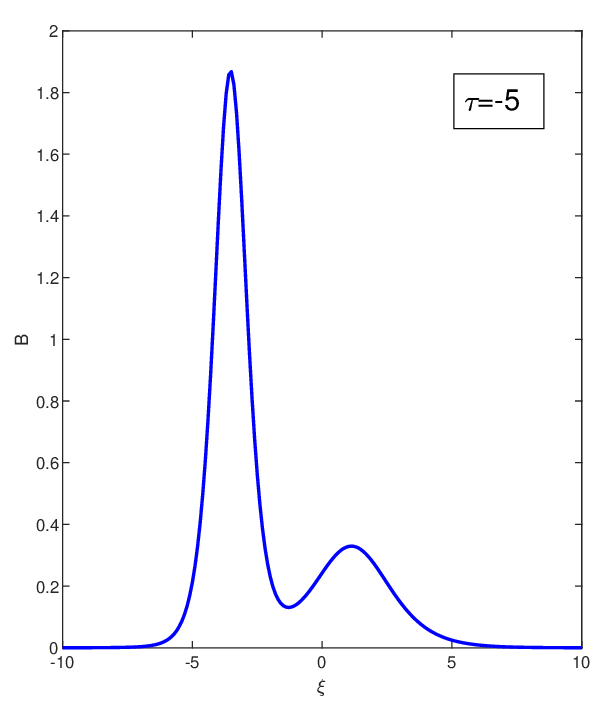}
     \caption{}
     \label{2soliton-b}
 \end{subfigure}
 \hfill
 \begin{subfigure}[t]{0.19\textwidth}
     \centering
     \includegraphics[width=\textwidth,height=2 in]{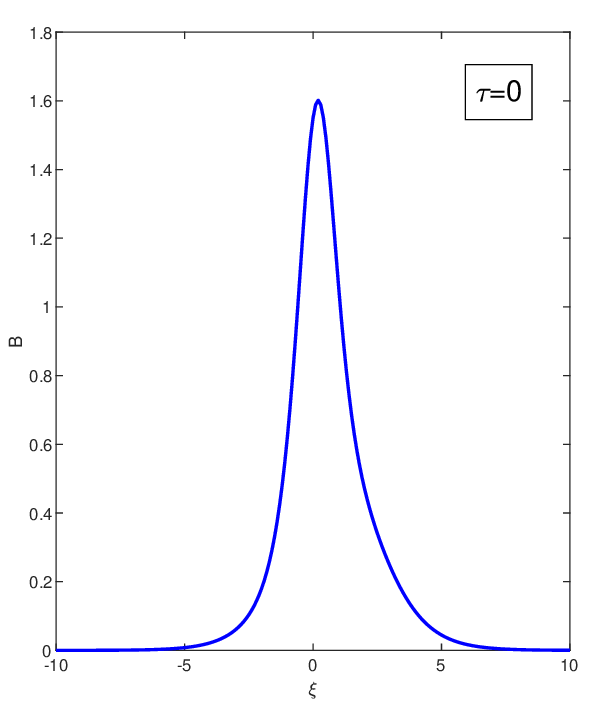}
     \caption{}
     \label{2soliton-c}
 \end{subfigure}
 \hfill
 \begin{subfigure}[t]{0.19\textwidth}
     \centering
     \includegraphics[width=\textwidth,height=2.03 in]{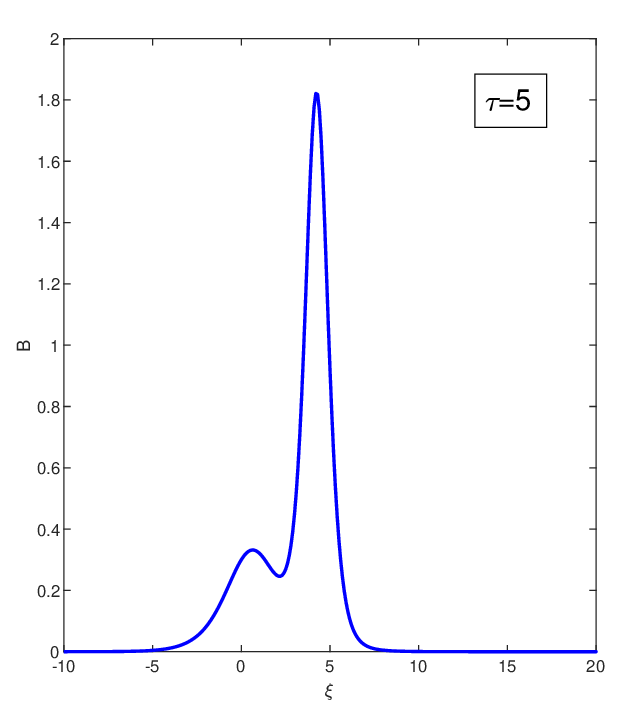}
     \caption{}
     \label{2soliton-d}
 \end{subfigure}
 \hfill
 \begin{subfigure}[t]{0.19\textwidth}
     \centering
     \includegraphics[width=\textwidth,height=1.98 in]{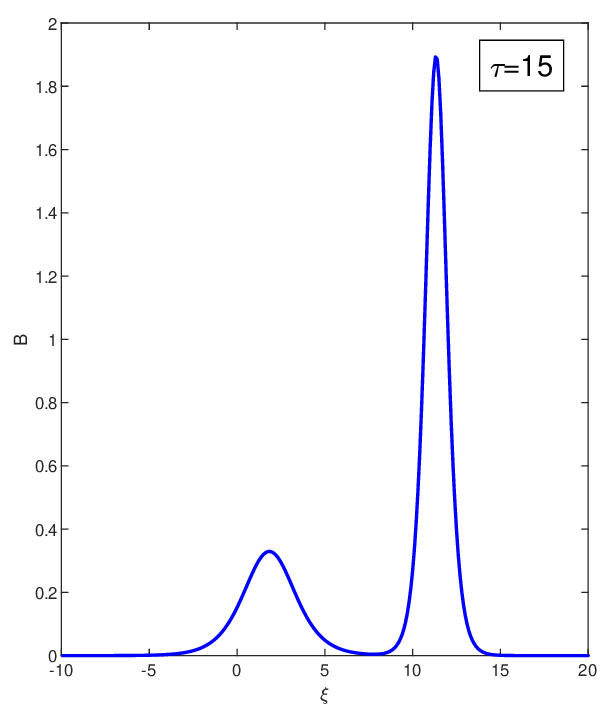}
     \caption{}
     \label{2soliton-e}
 \end{subfigure}
	\caption{\justifying Two-soliton profiles $B ~\text{vs}~\xi$ are displayed [see Eq. \eqref{eq-2soliton-solu}] for different values of $\tau$ showing the two solitons with different amplitudes collide and exchange energy and phase. For $\tau<0$ [see subplots (a) and (b)], the soliton with higher amplitude stays behind the lesser one.  At $\tau=0$ [see subplot (c)] two solitons collide and become one-soliton, after that, the soliton with higher amplitude moves faster than the lesser one [see subplots (d) and (e)]. Here, $(\Omega_0,\theta,C_g, C_c)=(0.7, \pi/4, 0.2, 0.4)$, $k_1=0.5$ and $k_2=1.2$. The other parameters remain unchanged as in Fig. \ref{fig-dispersion}.}
	\label{Fig-2soliton}
\end{figure*}
%%%%%%%%%%%%%%%%%%%%%%%%%%%%%%%%%%%
\section{NONLINEAR ANALYSIS} \label{sec-nls}
The linear analysis in Sec. \ref{Sec-Linear} reveals that the instability or damping of the magnetosonic waves is caused by the contribution of cosmic ray diffusivity ($\kappa$) and magnetic resistivity ($\eta$). The Coriolis force and the modified pressure, due to the rotational effects and cosmic ray effects, respectively, influenced the damping rate of the wave. The damping rate is estimated as $|\gamma| \sim (C_c^2 \kappa +\eta) k^4 / 2\omega_1^2$ in the linear perturbation. We investigate whether the linear perturbation can form nonlinear structures as they propagate and the nonlinear effect intervenes. Further, in the linear regime, perturbations are assumed to be very small, allowing us to disregard any nonlinear effects. In this section, we lighten the condition of small perturbation limits and intervene in nonlinear effects for which the linear theory is no longer valid, leading to the perturbation excitation of shocks, solitons, and rogue waves. To this end, we will derive evolution equations (KdVB equation in Subsection \ref{KdVB-Eq-Derivation} and the NLS equation in Section \ref{Sec-MI}) for such nonlinear waves in the preceding sections and examine their behaviour in different parametric regimes.

\subsection{KdVB Equation Derivation} \label{KdVB-Eq-Derivation}
In this subsection, we are interested in the non-linear evolution of progressive shocks of the one-dimensional plane in a frame that moves along the $x$-axis with the wave phase velocity. For this purpose, we derive an evolution equation for small-amplitude shocks and study their characteristics in physical parameter space. To employ the standard reductive perturbation technique, we stretch the space and time coordinates,
\begin{equation}
\label{stretched-co-or}
    \xi=\epsilon^{1/2} (x-\lambda t), \tau=\epsilon^{3/2} t,
\end{equation}
where $\lambda$ is the wave's phase velocity and the parameter $\epsilon$ measures the weakness of the nonlinearity. We expand the dependent variables in terms of $\epsilon$ as \cite{hager2023magnetosonic}
\begin{gather}
   \nonumber \rho=1+ \epsilon \rho_1 +\epsilon^2 \rho_2 +\dots,\\ 
   \nonumber v_x=0+ \epsilon v_{x1} +\epsilon^2 v_{x2} +\dots, \\
  \nonumber  v_z=0+ \epsilon v_{z1} +\epsilon^2 v_{z2} +\dots, \\
   \label{expansion} B=1+ \epsilon B_{1} +\epsilon^2 B_{2} +\dots, \\
   \nonumber P_g=P_{g0}+ \epsilon P_{g1} +\epsilon^2 P_{g2} +\dots, \\
   \nonumber P_c=P_{c0}+ \epsilon P_{c1} +\epsilon^2 P_{c2} +\dots,     
\end{gather}
while $v_y$ is expressed as
\begin{equation}
\label{expansion-vy}
    v_y=0+ \epsilon^{3/2} v_{y1} +\epsilon^{5/2} v_{y2} +\dots.
\end{equation}
Linear analysis in Sec. \ref{Sec-Linear} reveals that wave damping occurs due to the effect of $\eta$ and $\kappa$. Thus, we assume that $\eta \sim \eta_0 \epsilon^{1/2}$ and $\kappa\sim \kappa_0 \epsilon^{1/2}$ to make the perturbation evolution equation consistent, where $\eta_0$ and $\kappa_0$ are of the order of unity. The parameters $\eta$ and $\kappa$ are expanded in the lowest order of $\epsilon$ to take into account their effects for the propagation of dissipative magnetosonic shock waves numerically, which are valid as resistive and diffusive terms in Eq. \eqref{eq-n-mag} and \eqref{eq-n-cosmic}, respectively, are almost linearly proportional to the number density.  Similar assumptions have been made previously in many experimental situations \cite{nakamura1999observation,nakamura2001observation}. For large values of $\eta_0$ and $\kappa_0$, one can choose a higher order of $\epsilon$ (i.e., $\eta \sim \epsilon \eta_0$, $\kappa \sim \epsilon \kappa_0$) and in those cases sharp rising shock waves are noticed compared to oscillatory and monotonic shock waves (for $\eta \sim \epsilon^{1/2} \eta_0$, $\kappa \sim \epsilon^{1/2} \kappa_0$).

Inserting the stretched coordinates \eqref{stretched-co-or}, the expressions \eqref{expansion} and \eqref{expansion-vy} into the Eqs. \eqref{eq-n-con} to \eqref{eq-n-cosmic}, we identify the different powers of $\epsilon$. From the lowest order of $\epsilon$, we obtain the following expressions:
\begin{eqnarray}
\label{eq-1st-oreder-quan}
    v_{x1}=\lambda \rho_1 =\lambda B_1, P_{g1}=C_{g}^2 B_1, v_{z1}=\lambda \cot \theta B_1, \nonumber \\
    P_{c1}=C_{c}^2 B_1, v_{y1}=\frac{\lambda^2}{2 \Omega_0} \frac{\cot \theta}{\sin \theta} \frac{\partial B_1}{\partial \xi}, 
\end{eqnarray}
together with the compatible condition,
\begin{equation} \label{lambda_expression}
    \lambda^2=(1+C_{g}^2 +C_{c}^2) \times \frac{1}{(1+\cot^2 \theta)}.
\end{equation}
We obtained this cosmic ray modified phase velocity due to the interaction of ionized gas with cosmic rays. In this expression of the phase velocity \eqref{lambda_expression}, if one neglects the effects of the parameters related to modification of pressure due to  cosmic ray interaction, the expression agrees with the phase velocity as obtained in Ref. \cite{hager2023magnetosonic} in absence of Zeeman energy. Next, we obtain a set of equations in the next order of $\epsilon$. Eliminating those perturbed quantities from the equations and using the results given in Eq. \eqref{eq-1st-oreder-quan}, we derive the following KdVB equation, 
\begin{equation}
\label{eq-kdvb}
    \frac{\partial B}{\partial \tau} +Q B \frac{\partial B}{\partial \xi} +R \frac{\partial^3 B}{\partial \xi^3} =S \frac{\partial^2 B}{\partial \xi^2},
\end{equation}
where the first-order magnetic field perturbation  $B_1$ is replaced by $B$. The coefficient of nonlinearity $Q$, the coefficient of dispersion $R$ and the coefficient of dissipation $S$ are obtained as
\begin{eqnarray}
    & Q=\frac{\lambda \left(3+C_{g}^2+C_{c}^2 +\gamma_g C_{g}^2 +\gamma_c C_{c}^2 \right)}{\left(1+C_{g}^2+C_{c}^2+\lambda^2+\lambda^2\cot^2 \theta\right)},\\
    & R=\frac{\lambda^5 \cot^2 \theta}{4 \Omega_0^2 \sin^2 \theta} \times \frac{1}{\left(1+C_{g}^2+C_{c}^2+\lambda^2+\lambda^2\cot^2 \theta\right)}, \\
    & S=\frac{\left(\kappa_0 C_{c}^2 +\eta_0\right)}{\left(1+C_{g}^2+C_{c}^2+\lambda^2+\lambda^2\cot^2 \theta\right)}.
\end{eqnarray}
Here, all the coefficients $Q$, $R$ and $S$ are modified due to the presence of the parameters $C_g$, $C_c$ and $\theta$ related to the effects of thermal pressure, cosmic ray pressure and axis of rotation, respectively. However, in addition to these parameters, the diffusivity of cosmic rays ($\kappa_0$) and the magnetic resistivity ($\eta_0$) affect the dissipation coefficient $S$ whereas the rotation frequency ($\Omega_0$) affects the dispersion coefficient $R$.
In the absence of cosmic ray diffusivity ($\kappa_0=0$) and magnetic resistivity ($\eta_0=0$),  Eq. \eqref{eq-kdvb} reduces to the well-known KdV equation, and the solution can be obtained in terms of a solitary wave pulse. In other words, the parameters $\kappa_0$ and $\eta_0$ in the present model yield the formation of shock structures. In the limiting case, $\theta=\pi/2$, the dispersion coefficient vanishes, consequently Eq. \eqref{eq-kdvb} reduces to the Burgers equation having a stationary shock profile. Now, we move on to find various types of soliton and shock wave solutions from Eq. \eqref{eq-kdvb} for different cases.

%%%%%%%%%%%%%%%%%%%%%%
\subsection{Soliton Solutions} \label{Sec:Soliton Solutions}
In the turbulence zones of ISM of spiral galaxies, such as galaxy centre, star-forming regions, and supernova remnants, generally the magnetic Reynold number is high, and the magnetic fields follow the ``frozen-in-field'' condition on large scales. In such regions, magnetic fields can sustain over a long period of time and reduce cosmic ray diffusive ability by trapping them for longer periods. Therefore, it is conventional to consider the cases of weak magnetic resistivity ($\eta_0$) and cosmic ray diffusivity ($\kappa_0$). Neglecting the effect of $\eta_0$ and $\kappa_0$, one can obtain the following KdV equation from Eq. \eqref{eq-kdvb}, as
\begin{equation}
\label{eq-kdv}
    \frac{\partial B}{\partial \tau} +Q B \frac{\partial B}{\partial \xi} +R \frac{\partial^3 B}{\partial \xi^3} =0,
\end{equation}
which admits the well-known one-soliton solution given by,
\begin{equation}
    \label{eq-1soliton-solu}
    B=\frac{3v_0}{Q} \sech^2 \left[ \sqrt{\frac{v_0}{4R}} \zeta \right]
\end{equation}
where $\zeta=\xi-v_0 \tau$ and $v_0$ is the speed of the travelling wave. The typical profiles of one-soliton (solitary wave pulse) are plotted in Fig. \ref{Fig-solitary-profile} for different values of plasma parameters $\Omega_0$, $\theta$ (in the subplot (a)) and $C_g$, $C_c$ (in the subplot (b)), respectively. The width of the wave is found to become narrower, however, the amplitude of the wave remains unchanged as the rotating frequency $\Omega_0$ increases. But, if the angle of rotation $\theta$ increases, the amplitude and width of the solitary wave pulse decrease. Subplot (b) shows that a small increase in thermal pressure ($C_g$) and cosmic ray pressure ($C_c$) decrease the wave amplitude, but the width of the pulse becomes wider.

Now, we describe the transmission and interactions of two-soliton solutions for the KdV equation employing Hirota's bilinear method\cite{hirota1971exact}.
Recalling the variables $\xi$ by $\Tilde{\xi} B^{1/3}$, $B$ by $6\Tilde{B}{Q^{*}}^{-1} R^{1/3}$ and $\tau$ by $\Tilde{\tau}$, we rewrite the KdV equation as
\begin{equation}
    \label{eq-kdv-standard}
    \frac{\partial \Tilde{B}}{\partial \Tilde{\tau}} -6 \Tilde{B} \frac{\partial \Tilde{B}}{\partial \Tilde{\xi}} + \frac{\partial^3 \Tilde{B}}{\partial \Tilde{\xi^3}} =0.
\end{equation}
where $Q^{*}=-Q$. Assuming the transformation $\Tilde{B}=-2 \left(\log f\right)_{\Tilde{\xi}\Tilde{\xi}}$, we obtain the Hirota bilinear form of Eq. \eqref{eq-kdv-standard} as
\begin{equation}
    \label{eq-hirota-form}
    \left(D_{\Tilde{\xi}} D_{\Tilde{\tau}} +D_{\Tilde{\xi}}^4\right) \left\{ f\cdot f\right\}=0,
\end{equation}
where $f=1+\epsilon\left(e^{\Tilde{\theta}_{1}} + e^{\Tilde{\theta}_{2}}\right) + \epsilon^2 f_2$, $\Tilde{\theta}_{i}=k_i \Tilde{\xi} +\Tilde{\omega}_{i} \Tilde{\tau}$ ($i=1,2$) and $D$ represents Hirota's operator. Collecting various powers of $\epsilon$ from Eq. \eqref{eq-hirota-form}, we obtain $\Tilde{\omega}_i=-k_i^3$ and $f_2=a_{12} e^{\Tilde{\theta}_{1}+\Tilde{\theta}_{2}}$, with phase shift $a_{12}=\frac{(k_1-k_2)^2}{(k_1+k_2)^2}$. 
Finally, we have the two-soliton solution of Eq. \eqref{eq-kdv} obtained as
\begin{widetext}
 \begin{equation}
    \label{eq-2soliton-solu}
    B=-12 \frac{R^{1/3}}{Q^{*}} \frac{k_1^2 e^{\theta_1}+k_2^2 e^{\theta_2}+a_{12} e^{\theta_1+\theta_2} (k_2^2 e^{\theta_1}+k_1^2 e^{\theta_2}) +2(k_1-k_2)^2 e^{\theta_1+\theta_2}}{(1+e^{\theta_1}+e^{\theta_2}+a_{12} e^{\theta_{1}+\theta_{2}})^2},
\end{equation},   
\end{widetext}
where $\theta_i=\frac{k_i}{R^{1/3}} \xi -k_i^3\tau$ ($i=1,2$) and $\Tilde{B}$ is replaced by $\Tilde{B}=\frac{Q^{*} B}{6 R^{1/3}}$. As $\tau \gg 1$ the obtained solution  \eqref{eq-2soliton-solu} converts to the following superposed two individual single-soliton solutions, gradually, as follows\cite{drazin1989solitons}
\begin{equation}
    \label{eq-2soli-solu-superposed}
    B_{12}\sim \sum_{i=1}^{2} A_i \sech{\left[ \frac{k_i}{2R^{1/3}} (\xi -k_i^2 R^{1/3} \tau +\Delta_i)\right]},
\end{equation}
where $A_i= \frac{3 R^{1/3} k_i^2}{2 Q^{*}}$ ($i=1,2$) is the amplitude and the phase shifts of the solitons due to interactions are $\Delta_{1,2}=\pm  \frac{2 R^{1/3} \log|\frac{k_2-k_1}{k_2+k_1}|}{k_{1,2}}$.

 Numerically, we demonstrate the temporal evolution and phase shift of the two-soliton solution of the KdV equation by the interaction of two solitons. In Fig. \ref{Fig-2soliton}, the time evaluation of two solitons interactions $B$ vs $\xi$ have been plotted for different values of $\tau$. One can observe that two solitons with different amplitudes collide and exchange energy and phase. The shape and size of both solitons may alter during collision. Fig. \ref{Fig-2soliton} also depicts that for $\tau<0$, the soliton with higher amplitude stays behind the lesser one and as $\tau$ increases, the two solitons get closer. Subsequently, two solitons collide at $\tau=0$ and become one-soliton, after that, the soliton with higher amplitude moves faster than the lesser one, keeping the latter one behind. The temporal interaction of two solitons represents the mechanism for transporting energy and momentum in the ISM of spiral galaxies. Interaction between two solitons can lead to a localized amplification of energy and magnetic field. We noticed that when we changed the parametric values of the plasma model for the two-soliton solution, the characteristics framework stayed consistent with what we saw for the one-soliton solution.

%%%%%%%%%%%%%%%%%%%%%%%%%%%%%%%%%%%%%%%%%%%%%%%%

\begin{figure*}
	\centering
 \begin{subfigure}[t]{0.49\textwidth}
     \centering
     \includegraphics[width=\textwidth,height=2.5 in]{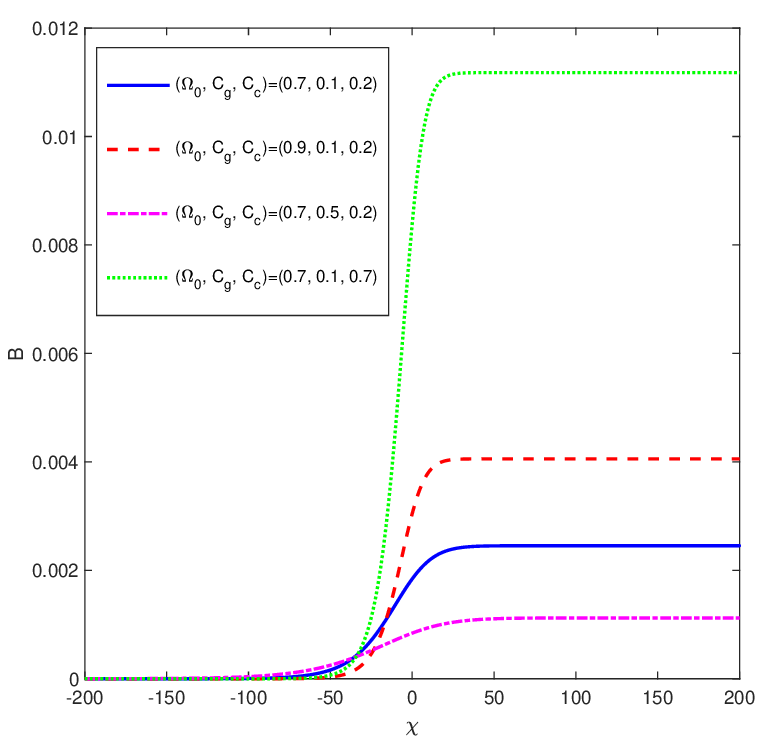}
     \caption{}
     \label{Fig-shock-profile-a}
 \end{subfigure}
 \hfill
 \begin{subfigure}[t]{0.49\textwidth}
     \centering
     \includegraphics[width=\textwidth,height=2.5 in]{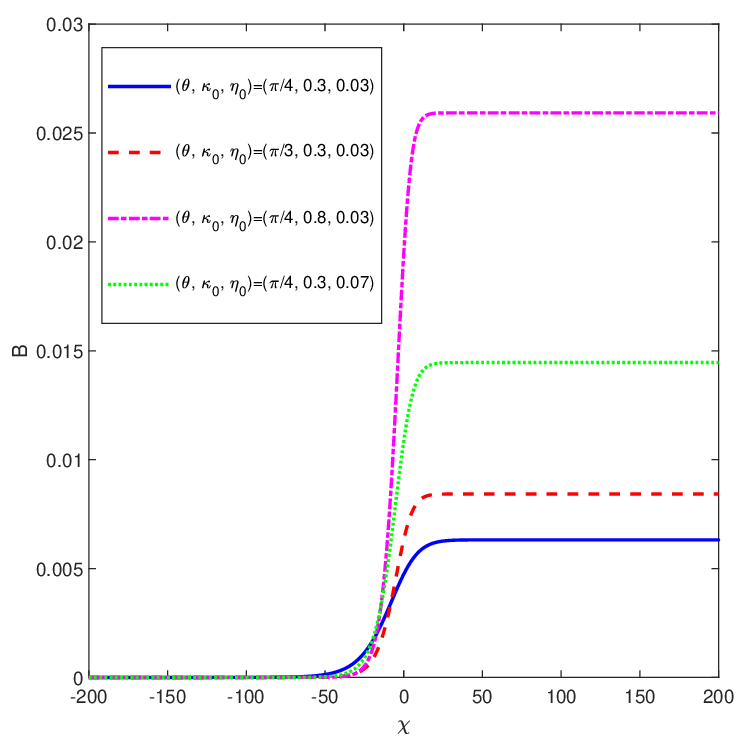}
     \caption{}
     \label{Fig-shock-profile-b}
 \end{subfigure}
	\caption{\justifying The magnetosonic shock profiles $B ~\text{vs}~\chi$ are plotted  [see Eq. \eqref{eq-kdvb-sol}] for different parametric values as mentioned in the legends. The amplitude is shifted to higher values and the width becomes narrower due to the increasing effect of rotational frequency ($\Omega_0$), angle of rotation ($\theta$), cosmic ray diffusivity ($\eta_0$), magnetic resistivity ($\kappa_0$) and the parameter $C_c$ related to cosmic pressure. However, the amplitude is shifted to lower values and the width becomes wider for increasing values of the parameter $C_g$ related to cosmic ray pressure. Here, ($\theta$, $\kappa$, $\eta$)= ($\pi/3$, 0.3, 0.03) and $(\Omega_0, C_g, C_c) = (0.9, 0.2, 0.4)$ for the subplots (a) and (b),
respectively. The other parameters remain unchanged as in Fig. \ref{fig-dispersion}.}
	\label{Fig-shock-profile}
\end{figure*}
%%%%%%%%%%%%%%%%%%%%%%%%%%%
\begin{figure*}
	\centering
 \begin{subfigure}[t]{0.49\textwidth}
    \centering
     \includegraphics[width=\textwidth,height=2.2 in]{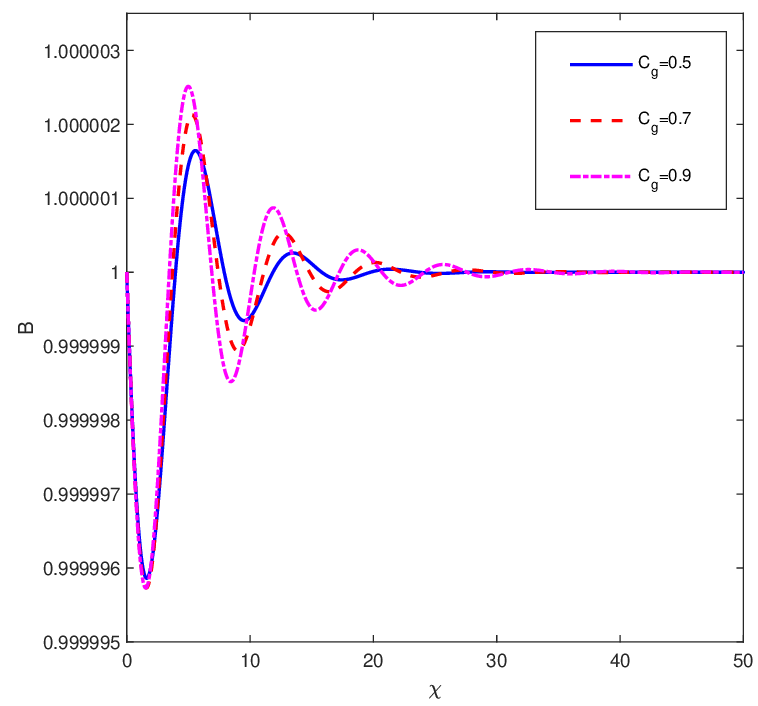}
     \caption{}
     \label{Fig-Monotonic-shocks-a}
 \end{subfigure}
 \hfill
\begin{subfigure}[t]{0.49\textwidth}
     \centering
     \includegraphics[width=\textwidth,height=2.2 in]{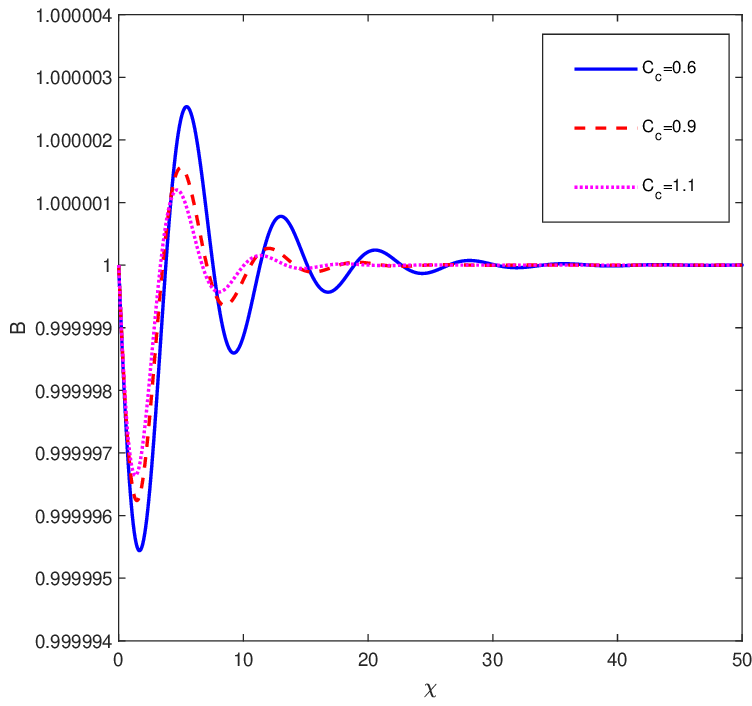}
     \caption{}
     \label{Fig-Monotonic-shocks-b}
 \end{subfigure}
 
 \begin{subfigure}[t]{0.49\textwidth}
     \centering
     \includegraphics[width=\textwidth,height=2.2 in]{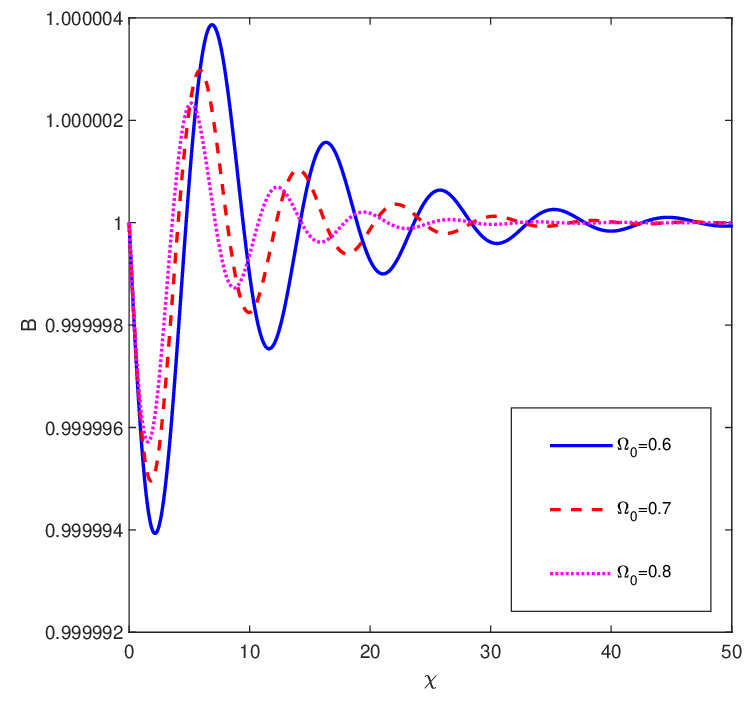}
     \caption{}
     \label{Fig-Monotonic-shocks-c}
 \end{subfigure}
 \hfill
 \begin{subfigure}[t]{0.49\textwidth}
     \centering
     \includegraphics[width=\textwidth,height=2.2 in]{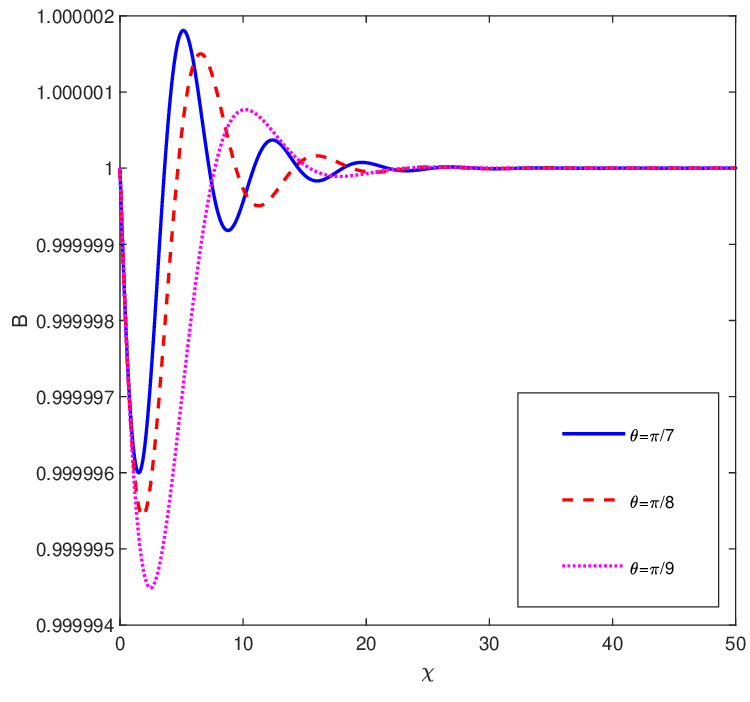}
     \caption{}
     \label{Fig-Monotonic-shocks-d}
 \end{subfigure}
 
 \begin{subfigure}[t]{0.49\textwidth}
     \centering
     \includegraphics[width=\textwidth,height=2.2 in]{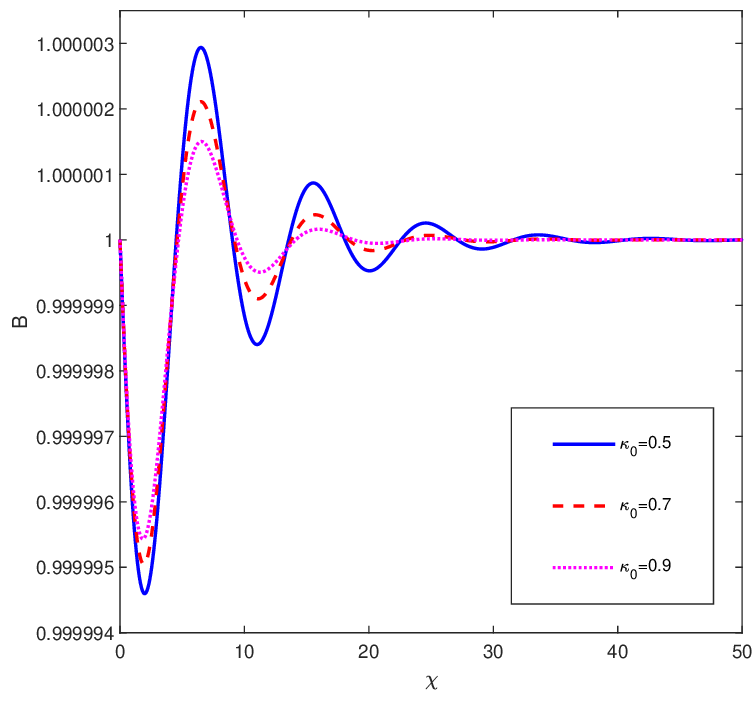}
     \caption{}
     \label{Fig-Monotonic-shocks-e}
 \end{subfigure}
 \hfill
 \begin{subfigure}[t]{0.49\textwidth}
     \centering
     \includegraphics[width=\textwidth,height=2.2 in]{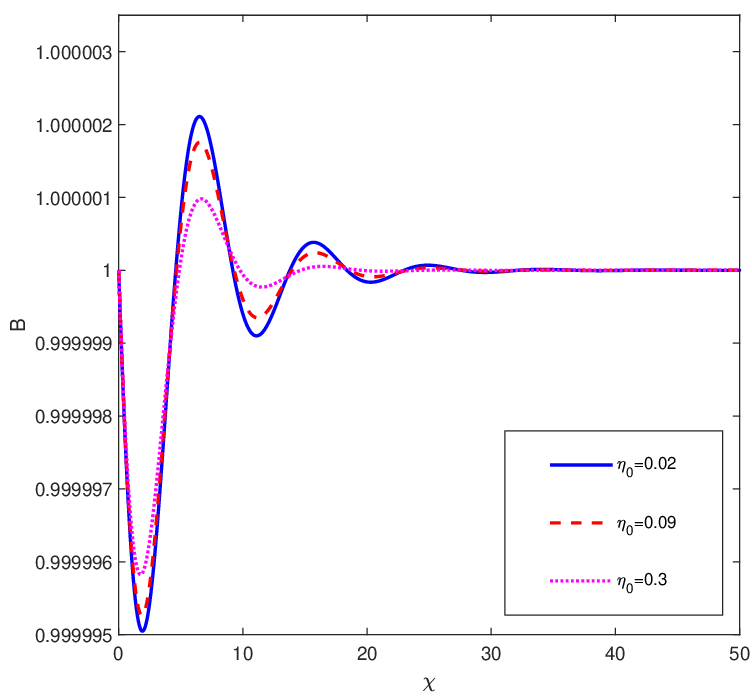}
     \caption{}
     \label{Fig-Monotonic-shocks-f}
 \end{subfigure}
% \hfill
        \caption{\justifying The magnetosonic oscillatory shock profiles $B~\text{vs}~\chi$ are plotted [see Eq. \eqref{eq-sym-ode}] for different values of the parameters as mentioned in the legends. The strength and width of the oscillatory shock grow as the parameter  $C_g$ related to thermal pressure and the angle of rotation ($\theta$)  increases, whereas they become smaller and tend to damp faster as the parameter $C_c$ related to cosmic pressure,  rotational frequency ($\Omega_0$), cosmic ray diffusivity ($\kappa_0$) and magnetic resistivity  ($\eta_0$)  increase. The parameters in the subplots, except the variation parameter, are fixed as $(C_g, C_c, \Omega_0, \theta, \kappa_0, \eta_0)=(0.7, 0.6, 0.8, \pi/7, 0.9, 0.02)$.}
	\label{Fig-Monotonic-shocks}
\end{figure*} 
%%%%%%%%%%%%%%%%%%%%%%%%%%%%%%%%%%%%%%%%%%%%%%%%
\subsection{Shock Wave Solutions}
After that, we discuss the case in which the effects of the parameter $\kappa_0$ and $\eta_0$ are considerable, and consequently, the dissipation coefficient $S$ plays a significant role in the evaluation of the shock wave solution of the KdVB equation \eqref{eq-kdvb}. The dispersion coefficient $R$ disappears for $\theta=\pi/2$ and, as a result, Eq. \eqref{eq-kdvb} reduces to the Burgers equation with a stationary shock solution as\cite{hossen2017electrostatic},
\begin{equation}
    B(\xi, \tau)=\frac{2U_0}{Q} \left[ 1-\tanh \left\{ \frac{U_0}{2S} (\xi-U_0 \tau) \right\} \right],
\end{equation}
where $U_0$ is the speed of the wave. For nonzero values of $R$, we find an analytic solution of \eqref{eq-kdvb} by employing Bernoulli's method \cite{hager2023magnetosonic} and do a numerical simulation to study the behaviours of monotonic and oscillatory shock wave pulses, respectively.

%%%%%%%%%%%%%%%%%%%%%%%%%%%%%%%%%%%%%%%%%%%%%%%%%%%%%%
  \begin{figure*}
	\centering
 \begin{subfigure}[t]{0.49\textwidth}
     \centering
     \includegraphics[width=\textwidth,height=2.5 in]{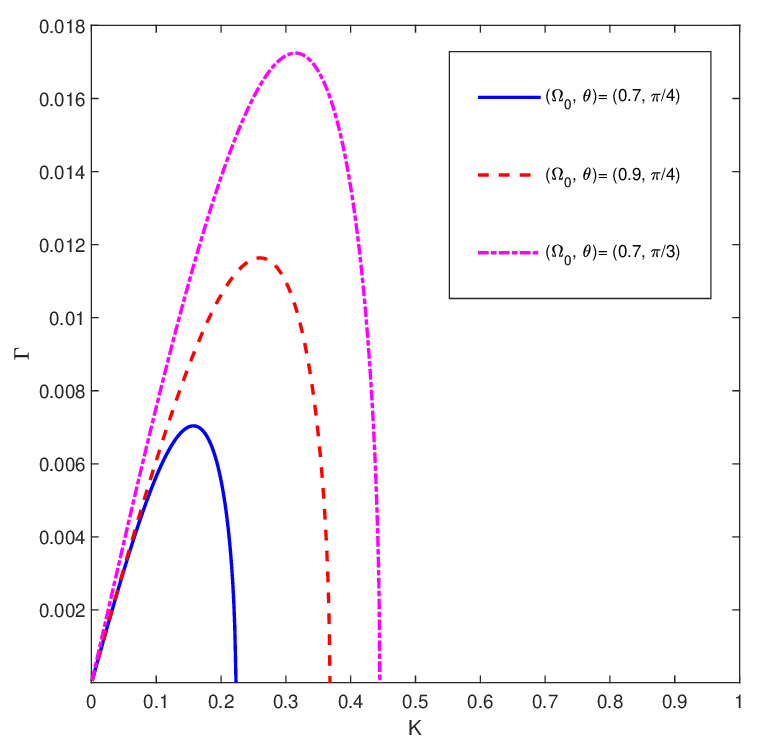}
     \caption{}
     \label{Growth-rate-fig-a}
 \end{subfigure}
 \hfill
 \begin{subfigure}[t]{0.49\textwidth}
     \centering
     \includegraphics[width=\textwidth,height=2.5 in]{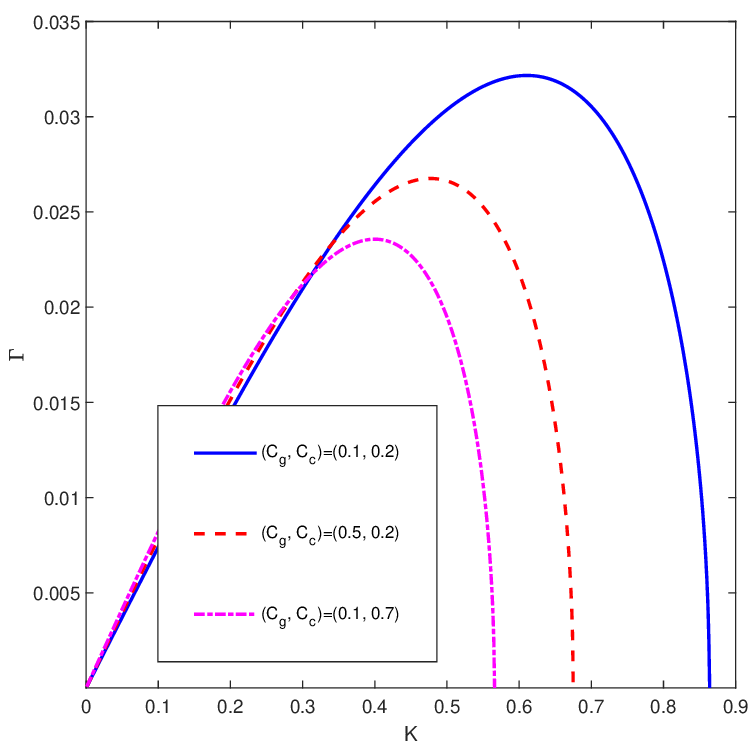}
     \caption{}
     \label{Growth-rate-fig-b}
 \end{subfigure}
	\caption{\justifying The development of the instability growth rate $\Gamma$ [see Eq. \eqref{growth-rate}] against the modulated wavenumber $K$ is shown for different values of parameters as mentioned in the legends. As the modulated wave number $K$ rises, the growth rate increases until it hits a critical growth rate, say $\Gamma_c$. For progressively higher values of $K$, it sharply declines after reaching the critical value $\Gamma=\Gamma_c$. Here, $B_0=0.04$ and other parameters remain unchanged as in Fig. \ref{Fig-solitary-profile}.}
	\label{Growth-rate-fig}
\end{figure*}
%%%%%%%%%%%%%%%%%%%%%%%%%%%%%%%%%%%%%%%%%%%%%%%%%%%%%
%%%%%%%%%%%%%%%%%%%%%%%%%%%%%%%%
\begin{figure*}
	 \centering
 \begin{subfigure}[t]{0.49\textwidth}
     \centering
     \includegraphics[width=\textwidth,height=2.5 in]{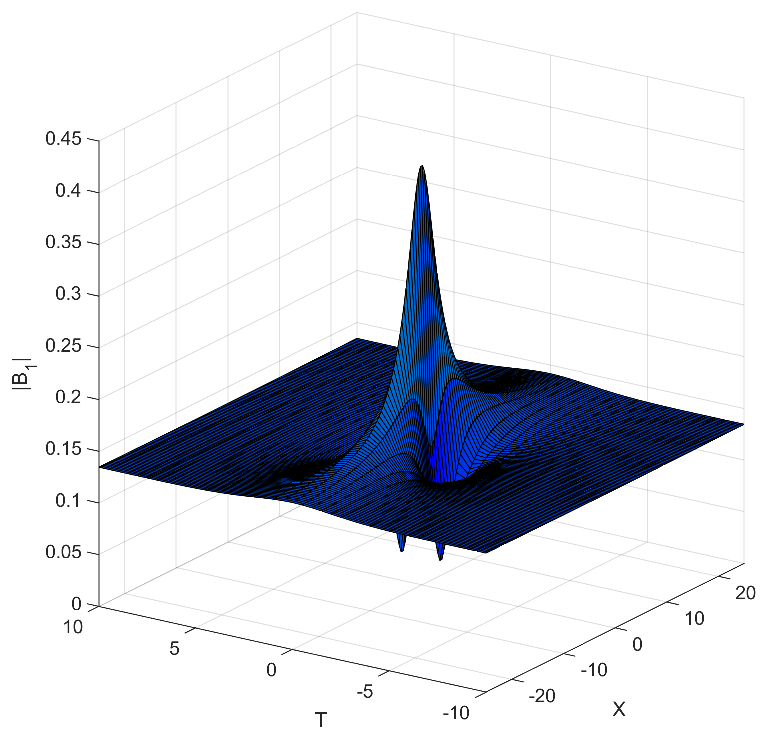}
     \caption{}
     \label{fig-3d-rogue-wave-a}
 \end{subfigure}
 \hfill
  \begin{subfigure}[t]{0.49\textwidth}
     \centering
     \includegraphics[width=\textwidth,height=2.5 in]{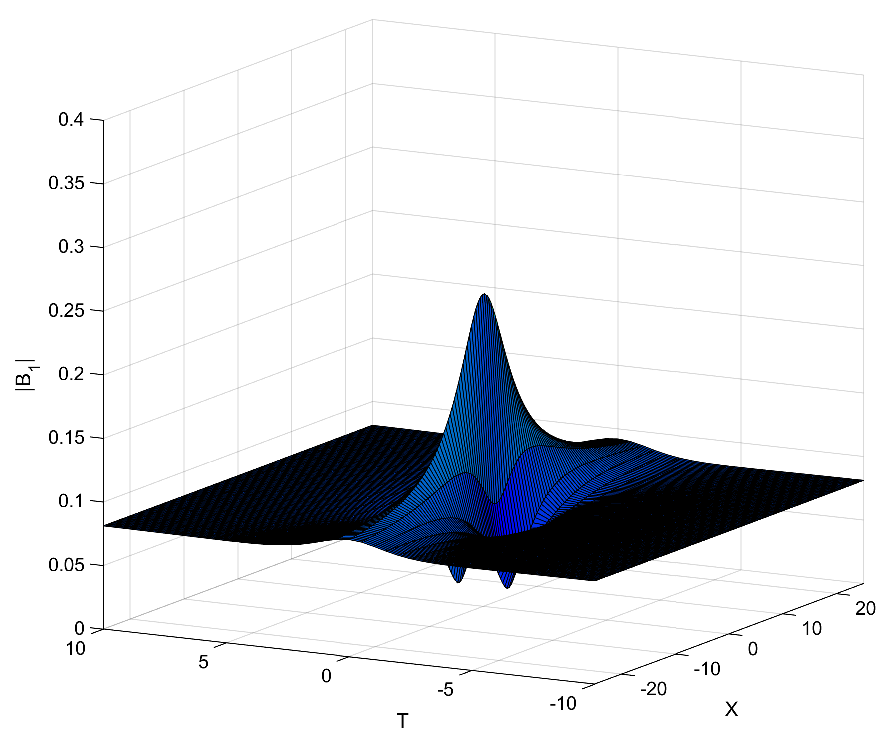}
     \caption{}
     \label{fig-3d-rogue-wave-b}
 \end{subfigure}
	\caption{\justifying The three-dimensional first-order rogue wave profiles $|B_1|$  are displayed [see Eq. \eqref{eq-1st-order-rogue}] for two different values of the rotational frequency, $\Omega_0=0.7$ and $\Omega_0=0.9$ in the subplots (a) and (b), respectively. The increase in rotational frequency results in a decrease in the rogue wave's amplitude and width. The observation is quite natural, as the pulse cannot gain energy from the background wave. Here, $(C_g, C_c,\theta)=(0.1, 0.7, \pi/3)$ and all other parameters  remain unchanged  as in Fig. \ref{Fig-solitary-profile}.}
	\label{fig-3d-rogue-wave}
\end{figure*} 
%%%%%%%%%%%%%%%%%%%%%%%%%%%%%%%%%%%

\subsubsection*{Analytic Solution}
Using the stationary wave frame $\chi=-(\xi-v \tau)$ moving with speed $v$ and integrating over the variable $\chi$, we obtain, 
\begin{equation}
\label{eq-ode-trav}
    R \frac{d^2 B}{d \chi^2}+S \frac{d B}{d \chi} +\frac{Q}{2} B^2-v B=0,
\end{equation}
where we use the boundary conditions $B, \frac{d B}{d \chi}, \frac{d^2 B}{d \chi^2} \rightarrow 0$ as $\chi \rightarrow\pm \infty$. Following the method, we obtain a travelling wave solution of  \eqref{eq-kdvb} as,
\begin{equation}
\label{eq-BG} 
    B(\chi)=a_0 + a_1 G(\chi) + a_2 G^2(\chi),
\end{equation}
where $G(\chi)=\sigma/2 \left\{1+\tanh{[(\sigma/2)\chi]}\right\}$ is the solution of the Bernoulli's equation,
\begin{equation}
    \frac{d G(\chi)}{d \chi}=\sigma G(\chi) -G^2(\chi). 
\end{equation}
and the unknown constants $a_i$ ($i=0,1,2$) and $\sigma$ are to be calculated later. Differentiating Eq. \eqref{eq-BG} up to second order with respect to $\chi$, we obtain,
\begin{equation}
\label{eq-dbdx}
    \frac{d B}{d \chi}= (a_1 +2 a_2 G) (\sigma G -G^2)
\end{equation}
and
\begin{equation}
\label{eq-d2bdx2}
    \frac{d^2 B}{d \chi^2}=[a_1 (\sigma-2G)+2 a_2 (2\sigma G-3 G^2)] (\sigma G-G^2).
\end{equation}
Plugging the expressions \eqref{eq-BG}, \eqref{eq-dbdx} and \eqref{eq-d2bdx2} into \eqref{eq-ode-trav} and collecting various terms of the same degree of $G$, we get,
\begin{equation}
    a_0= 2\frac{v}{Q}, a_1=0, a_2=-12 \frac{R}{Q}, \sigma=-\frac{S}{5R}
\end{equation}
and the speed of the moving frame $v=\frac{6S^2}{25 R}$. Therefore, the localized solution of the KdVB equation \eqref{eq-kdvb} is written as follows,
\begin{equation}
\label{eq-kdvb-sol}
B=\frac{3 S^2}{25 QR} \left[\sech^2{(\frac{S}{10R}\chi)}+2 \left\{ 1+\tanh{(\frac{S}{10R}\chi)}\right\}\right],
\end{equation}
which describes a composite form of solitary and Burger shock structure. In other words, the solution \eqref{eq-kdvb-sol} represents an admixture of  $\sech^2$ and  $\tanh$ type waveform. The graphical representations of such profiles have been displayed in Fig. \ref{Fig-shock-profile} for distinct parametric values, as stated in the caption of the corresponding figure. The figures reveal that, due to the increments in the values of the parameters $\Omega_0$, $\theta$, $\eta_0$ and $\kappa_0$, the amplitude is shifted to higher values, while the width becomes narrower for the monotonic shock wave pulses. Thus, cosmic ray diffusivity and magnetic resistivity play a key role in determining the nature and magnitude of the shock pulse in a dissipative magneto-rotating plasma under the effect of Coriolis force.  The effect of modified pressure by combining the cosmic ray pressure (relating parameter $C_c$) with thermal pressure (relating parameter $C_g$) has been observed from the subplot (a) of Fig. \ref{Fig-shock-profile}. It has been noticed that when $C_c$ is fixed, the amplitude is shifted to lower values and the width becomes wider for increasing values of $C_g$; however, the amplitude increases and the width becomes narrower for increasing values of $C_c$. Due to the effect of cosmic ray modified pressure, cosmic ray diffusivity, and magnetic resistivity, the dissipation term in Eq. \eqref{eq-kdvb} is dominant. Therefore, Fig. \ref{Fig-shock-profile} indicates that the energy dissipation caused by the cosmic ray effects leads to the temporal evaluation of shock wave profiles in the ISM of spiral galaxies. In other words, the ISM's shock wave structures are key mechanisms for accelerating cosmic rays.  

%%%%%%%%%%%%%%%%%%%%%%%%%%%%%%%%%%%%%
\subsubsection*{Numerical Solution}
Now, we numerically solve the Eq. \eqref{eq-kdvb} to look into further potential structures and their characteristics. To do so, we present Eq. \eqref{eq-kdvb} in the following coupled differential equation as
\begin{equation}
\label{eq-sym-ode}
    \begin{split}
        &\frac{dB}{d\chi}=Z,\\
        &\frac{dZ}{d\chi}=\frac{v}{R} B-\frac{Q}{2R} B^2-\frac{S}{R}Z,
    \end{split}
\end{equation}
which admits two fixed points, $(0,0)$ and $(2v/Q,0)$. Next, we employ the fourth-order Runge-Kutta method to solve Eq. \eqref{eq-kdvb} to study the effects of different parameters involved in the model, and the results obtained are displayed in Fig. \ref{Fig-Monotonic-shocks}. Subplots (a) and (b) show the effects of $C_g$ and $C_c$ on the oscillatory shock structures. We observe that as $C_g$ increases, the strength and width of the oscillatory shock grow, whereas they become smaller and tend to damp faster as $C_c$ increases. Subplots (c) and (d) illustrate how the rotational frequency $\Omega_0$ with the angle of rotation $\theta$ affects the oscillatory shock profiles. It is found that the oscillatory shock wave strength and width become smaller and tend to damp faster with increasing values of $\Omega_0$ with a fixed angle of rotation. On the other hand, the amplitude and width increase as $\theta$ increases for a fixed angle of rotation. Subplots (e) and (f) indicate that due to small increments in cosmic ray diffusivity $\kappa_0$ and magnetic resistivity $\eta_0$, the strength and width of oscillatory shock structures become smaller and tend to damp faster. The physical significance behind this phenomenon is that when the dissipative effects dominate the plasma environment, the monotone shock profile is formed, otherwise, the behaviour remains oscillatory.
%%%%%%%%%%%%%%%%%%%%%%%%%%%%%%%%%%

%%%%%%%%%%%%%%%%%%%%%%%%%%%%%%%%%%%%%%%%%%%%%%%%%%%%
\section{MODULATIONAL INSTABILITY AND ROGUE WAVE SOLUTIONS} \label{Sec-MI}
In this section, we are interested in studying the magnetosonic wavepackets' modulation instability (MI) and the characteristics of rogue wave solutions. In the limit of low frequency (i.e., when the carrier wave frequency is much smaller than the plasma frequency), neglecting the effect of magnetic resistivity and cosmic ray diffusivity due to high Reynold number and ``frozen-in-field'' condition in the turbulence zones of ISM of spiral galaxies as discussed in  Sec. \ref{Sec:Soliton Solutions}, we derive a nonlinear Schr\"{o}dinger (NLS) equation from Eq. \eqref{eq-kdvb} that describes the weakly nonlinear evolution of the modulated wave packet's envelope. To this end, we consider the solution of Eq. \eqref{eq-kdvb} in the form of a weakly modulated sinusoidal wave,
\begin{equation}
\label{expansion-b-nls}
B= \sum_{n=1}^{\infty} \epsilon^n \sum_{l=-\infty}^{\infty} B_{l}^{(n)}(X, T) \exp\left[il\left(k \xi-\omega \tau \right)\right].
\end{equation}
Here, $k$ and $\omega$ are carrier wave number and angular frequency, respectively. Further, we stretched the coordinates, $X$ and $T$ as
\begin{equation} 
\label{stretched-nls}
	\begin{split}
	X &=\epsilon\left(\xi-v_{g} \tau\right), \\
	T &=\epsilon^{2} \tau,	
	\end{split}
\end{equation}
where $v_g$ denotes the group velocity, which will be calculated later. All perturbed quantities are supposed to vary on the fast scales through the phase $(k \xi -\omega \tau)$ only, while the slow scales (X, T) enter the arguments of the $l$-th harmonic amplitude $B_l^n$. $B_l^{(n)}$ is kept equal to its complex conjugate $B_l^{(n)*}$ to ensure that $B$ remains real. The two partial differential operators $\partial / \partial \tau$ and $\partial / \partial \xi$  can be replace as follows,
\begin{equation} 
\label{operator-stretched}
	\begin{split}
	\frac{\partial}{\partial \tau}&\rightarrow \frac{\partial}{\partial \tau}-\epsilon v_g\frac{\partial}{\partial X} +\epsilon^{2}\frac{\partial}{\partial T}, \\
	\frac{\partial}{\partial \xi}&\rightarrow \frac{\partial}{\partial \xi} +\epsilon \frac{\partial}{\partial X}.	\\ \\
	\end{split}
\end{equation}
Substituting the expressions from \eqref{expansion-b-nls}-\eqref{operator-stretched}, into the Eq. \eqref{eq-kdvb}, we obtain,
\begin{widetext}
\begin{equation}
\label{eq-harmonic-order}
        \begin{split}
            -i\omega l B_{l}^{(n)}-v_g  \frac{\partial}{\partial T} B_{l}^{(n-1)}+\frac{\partial}{\partial T} B_{l}^{(n-2)} 
            + A \sum_{n=1'}^{\infty} \sum_{l'=-\infty}^{+\infty} \left( ilk {B}_{l-l'}^{(n-n')} B_{l'}^{(n')}  + {B}_{l-l'}^{(n-n'-1)} \frac{\partial}{\partial X } B_{l'}^{(n')}\right) \\
           + B\left( -i l^3 k^3 B_l^{(n)} -3 l^2 k^2 \frac{\partial {B}_{l}^{(n-1)}}{\partial X} +3ilk \frac{\partial^2 {B}_{l}^{(n-2)}}{\partial X^2} +\frac{\partial^3 {B}_{l}^{(n-3)}}{\partial X^3}\right)=0.
        \end{split}
    \end{equation}    
\end{widetext} 
In Eq. \eqref{eq-harmonic-order}, from the first order with the first harmonic (i.e. $n=1$, $l=1$), we obtain  the linear dispersion relation,
\begin{equation}
    \omega=-R k^3.
\end{equation}
From the second order with the first harmonic (i.e. $n=2$, $l=1$), we obtain the expression for group velocity as
 \begin{equation}
     v_g=3 R k^2=\frac{\partial \omega}{\partial k}.
 \end{equation}
Similarly, for $n=2$ and $l=2$, we obtain  $B_2^{(2)}=\frac{A}{6 R k^2}B_1^{(1)^2}$ and 
for $n=3$, $l=0$, we get $B_0^{(2)}=-\frac{Q}{v_g} |B_1^{(1)}|^2$.
Proceeding to the third order ($n=3$) with the first harmonic ($l=1$), we obtain an explicit compatibility condition leading to the NLS equation,
\begin{equation}
    	\label{nls_eq}
	i \frac{\partial B}{\partial T} + \frac{1}{2} M \frac{\partial^2 B}{\partial X^2} +N |B|^2 B=0,
\end{equation}
where $B_1^{(1)}$ is replaced by $B$ for simplicity. The dispersion and the nonlinear coefficients in Eq. \eqref{nls_eq} are  given by
\begin{equation} 
	M = 6 Rk
 \end{equation}
 \text{and}
 \begin{equation}
  N =\frac{Q^2}{6Rk},   
 \end{equation} 
respectively. Eq. \eqref{nls_eq} describes the nonlinear evolution of modulated magnetosonic wave packet envelopes, which can also be obtained directly by plugging the Eqs. \eqref{expansion-b-nls} and \eqref{operator-stretched} into Eqs. \eqref{eq-n-con}-\eqref{eq-n-cosmic}. In that case, the NLS represents an arbitrary frequency wave carrier with much more complicated expressions for $M$ and $N$.
However, in principle, the arbitrary frequency NLS equation should bring down to the Eq. \eqref{nls_eq} in the limit of low-frequency wave \cite{ruderman2008dynamics}. The study of such low-frequency waves and the modulation of sinusoidal waves, which describe various physical phenomena in plasma, is quite natural.

To study the MI, we consider a small perturbation $\delta B$ such that $B=B_0+\delta B\exp{i \Delta B}$. $B_0$ is the amplitude of the carrier wave such that $|B_{0}| \gg |\Delta B|$, $\Delta$ being the nonlinear frequency shift. Using the expansion of $B$ in Eq. \eqref{nls_eq}, we obtain,
\begin{equation}
  \Delta  =-Q |B_0|^2 
\end{equation}
and
\begin{equation}
  i \frac{\partial B}{\partial T} +\frac{1}{2} M \frac{\partial^2 B}{\partial X^2} +N	|B_0|^2 (\delta B+\delta B^{*})=0,  
\end{equation}
where $\delta B^{*}$ is the complex conjugate of $\delta B$. Assuming the amplitude perturbation varies as $\exp{(iKX-i \Tilde{\Omega} T)}$, we derive the nonlinear dispersion relation as follows,
\begin{equation}
    \label{eq-nonlinear-dispersion}
    \Tilde{\Omega}^2 = M^2 K^2(K^2- \frac{2 B_0 N}{M}).
\end{equation}
Here, $K$ and $\Tilde{\Omega}$ are modulated wave number and frequency, respectively. Based on the nonlinear analysis of the envelope dynamics, it is assumed that there are two kinds of envelope solitons, bright (unstable) and dark (stable). If $MN<0$, Eq. \eqref{eq-nonlinear-dispersion} possesses real solutions in terms of $\Tilde{\Omega}$ for all real values of $K$, consequently, the magnetosonic waves become dark (stable) envelope solitons in this region. However, in our case where $MN>0$, the region carries bright (unstable) envelope solitons in the presence of small external perturbation. In this region, there exists a critical modulated wave number $K_c=\sqrt{2N |B_0|^2/M}$ below which $\Tilde{\Omega}$ becomes imaginary, and MI takes place. For $MN>0$ and $K<K_c$, the growth rate of MI is specified as,
 \begin{equation}
 \label{growth-rate}
     \Gamma=|M| K^2 \sqrt{\frac{K_c^2}{K^2}-1}.
 \end{equation}
 The maximum growth rate is $\Gamma_{max}=N|B_0|^2$ at $K=K_c/\sqrt{2}$. The growth rate of instability depends on the coefficients $M$ and $N$ (which are further influenced by various plasma parameters), any changes in the plasma parameters certainly change the growth rate of instability. The coefficients $M$ and $N$ are modified due to the inclusion of cosmic ray pressure, which affects the growth of the instability. The graphs of the growth rate $\Gamma$ have been plotted against the modulated wave number $K$ in Fig. \ref{Growth-rate-fig} for different values of rotational frequency $\Omega_0$, angle of rotation $\theta$ (see subplot (a)), the parameter associated with thermal pressure $C_g$ and cosmic ray pressure $C_c$ (see subplot (b)), respectively. We observe that $\Gamma$ increases with the increasing value of $\Omega_0$ and $\theta$, however, it decreases as the values of $C_g$ and $C_c$ increase. It has been noticed that including cosmic ray pressure in the present model reduces the instability growth.  Furthermore, the plots in Fig. \ref{Growth-rate-fig} show that the growth rate increases with increasing modulated wave number $K$ until it reaches a critical growth rate (say, $\Gamma_c$), which varies for different plasma parameters. After reaching the critical valued growth rate, it intensely decreases for further higher values of $K$. In particular, one can find $\Gamma_c=0.017$ at $K=0.3$ in the subplot (a) of Fig. \ref{Growth-rate-fig} for $k=0.4$ and $\theta=\pi/3$.
 %%%%%%%%%%%%%%%%%%%%%%%%%%%%%%%%%%%%%
\begin{figure*}
	\centering
 \begin{subfigure}[t]{0.49\textwidth}
     \centering
     \includegraphics[width=\textwidth,height=2.5 in]{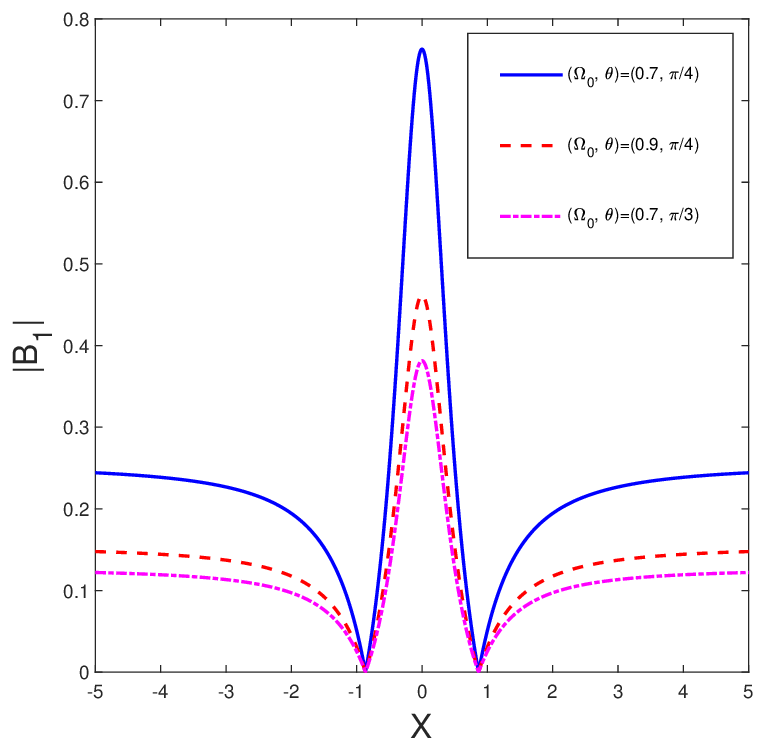}
     \caption{}
     \label{fig-1st-order-rogue-a}
 \end{subfigure}
 \hfill
 \begin{subfigure}[t]{0.49\textwidth}
     \centering
     \includegraphics[width=\textwidth,height=2.5 in]{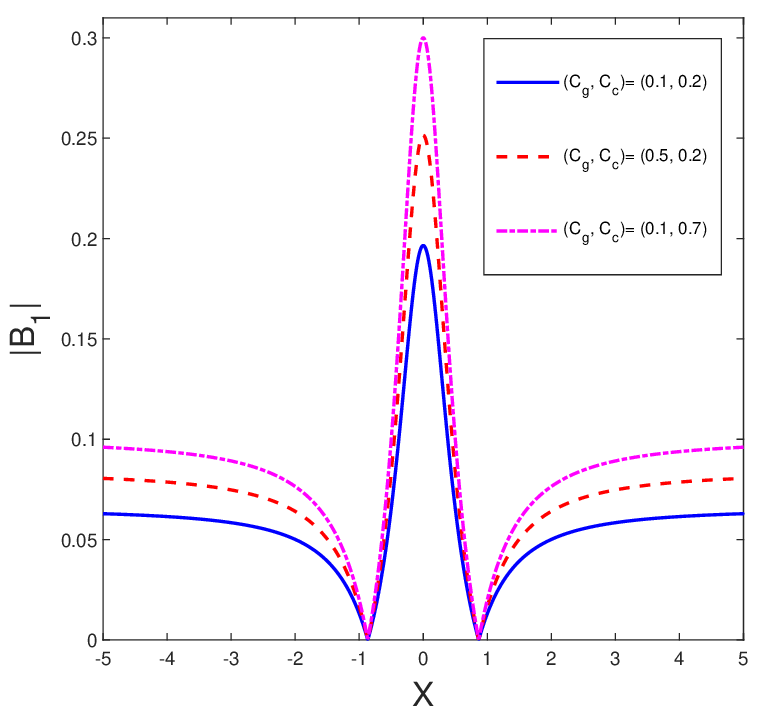}
     \caption{}
     \label{fig-1st-order-rogue-b}
 \end{subfigure}
    	\caption{\justifying The first-order rogue wave pulses $|B_1|$ vs X [see Eq. \eqref{eq-1st-order-rogue}] are plotted for different values of parameters as mentioned in the legends. The amplitude and width are found to be increased with the increasing value of the parameters $C_g$ and $C_c$ related to the thermal pressure and cosmic ray pressure, respectively. An increase in rotational frequency ($\Omega_0$) and/or angle of rotation ($\theta$)  leads to the contraction of the rogue wave pulse. Here, $K=0.4$ and all other parameters remain unchanged as in Fig. \ref{Fig-solitary-profile}.}
	\label{fig-1st-order-rogue}
\end{figure*}
%%%%%%%%%%%%%%%%%%%%%%%%%%%%%%%%
\begin{figure*}
	\centering
 \begin{subfigure}[t]{0.49\textwidth}
     \centering
     \includegraphics[width=\textwidth,height=2.5 in]{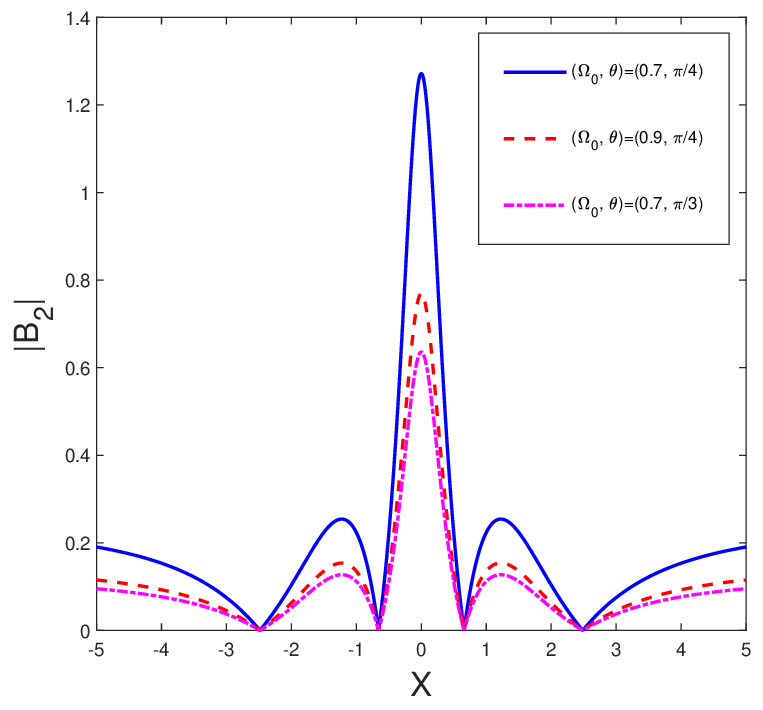}
     \caption{}
     \label{fig-2nd-order-rogue-a}
 \end{subfigure}
 \hfill
 \begin{subfigure}[t]{0.49\textwidth}
     \centering
     \includegraphics[width=\textwidth,height=2.5 in]{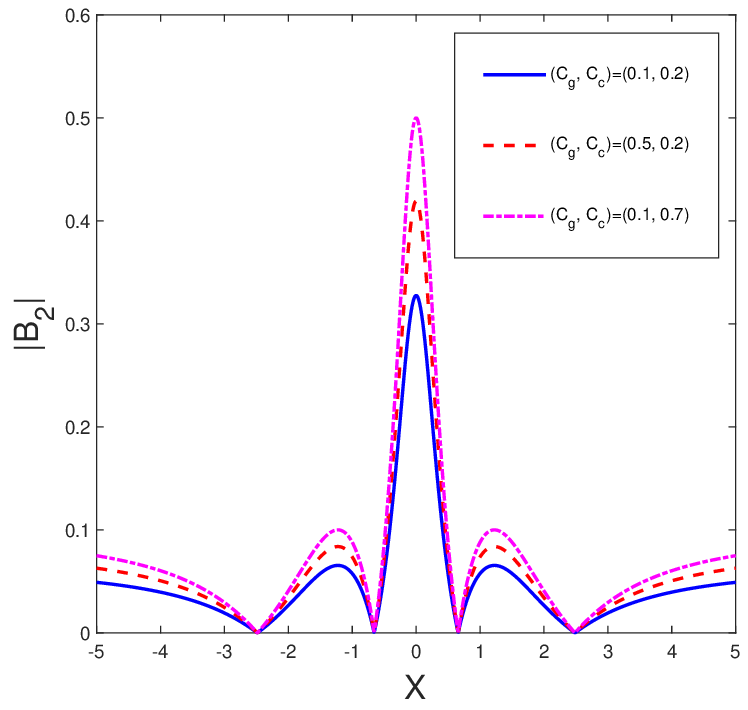}
     \caption{}
     \label{fig-2nd-order-rogue-b}
 \end{subfigure}
	\caption{\justifying The second-order rogue wave pulses $|B_2|$ vs X [see Eq. \eqref{eq-2nd-order-rogue}] are plotted for different values of parameters as mentioned in the legends. The amplitude of the second-order pulse is higher than that of first-order rogue wave pulses as observed in Fig. \ref{fig-1st-order-rogue}.  The effect of the parameters $\Omega_0$,  $\theta$, $C_g$ and $C_c$ are found to be similar as that of first-order pulses. Here, $K=0.4$ and all other parameters remain unchanged as in  Fig. \ref{Fig-solitary-profile}.}
	\label{fig-2nd-order-rogue}
\end{figure*}
%%%%%%%%%%%%%%%%%%%%%%%%%%%%%%%%%

\subsection{Rogue Wave Solutions}
The  MI of the magnetosonic wavepackets leads to the generation of freak waves. Generations of such waves in the region $MN>0$, are noticed to appear instantaneously and depart without any track down. Subsequently, the NLS Eq. \eqref{nls_eq} admits rational solutions localized in both space and time variables \cite{kharif2008rogue, moslem2011dust} as follows,
\begin{equation}
\label{eq-1st-order-rogue}
    B_1(X,T)= \sqrt{M/N} \left[ \frac{4(1+2iMT)}{1+4X^2+4 M^2 T^2}-1\right] \exp(iMT).
\end{equation}
The first-order rational rogue solution \eqref{eq-1st-order-rogue} reveals that a significant amount of magnetosonic wave energy is concentrated in a relatively small area in space. Typically, rogue waves are the envelopes of carrier waves having wavelengths smaller than those of the central region of the envelope. These waves have the unusual feature of not being periodic in both space and time. Combinations of multiple first-order rogue waves by nonlinear superpositions lead to higher-order rogue wave solutions having more complicated nonlinear forms and higher amplitude. The amplitude of such waves is generally four to five times the amplitude of the background wave. The existence of second-order rogue waves was experimentally observed by Chabchoub \textit{et al.} \cite{chabchoub2012super}  in surface water gravity waves. The exact second-order rogue wave solution  of the NLS equation \eqref{nls_eq} is given by \cite{akhmediev2009rogue},

\begin{equation}
\label{eq-2nd-order-rogue}
    B_2(X, T) =\sqrt{M/N} \left( 1 + \frac{M_2+iN_2}{O_2}\right) \exp(iM T),
\end{equation}\\
where 
\begin{equation}
    M_2=\frac{3}{8}-\frac{1}{2} X^4-\frac{3}{2} X^2-6 (P X T)^2 -10 (P T)^4-9 (PT)^2, \nonumber
\end{equation}
\begin{eqnarray}
    N_2=-P T \biggr[ -\frac{15}{4} +X^4 -3 X^2 + 4 (P X T)^2 + 4 (P T)^4  \nonumber\\ 
    +2 (P T)^2\biggr], \nonumber
\end{eqnarray}

\begin{eqnarray}
     O_2=\frac{3}{32} +\frac{1}{12} X^6+\frac{1}{8} X^4 +\frac{1}{2} X^4 (P T)^2+\frac{9}{16} X^2 +X^2(P T)^4 \nonumber \\
    -\frac{3}{2} (P X T)^2 +\frac{2}{3} (P T)^6+\frac{9}{2} (P T)^4+\frac{33}{8} (P T)^2. \nonumber
\end{eqnarray}

\begin{figure*}
	\centering
 \begin{subfigure}[t]{0.49\textwidth}
     \centering
     \includegraphics[width=\textwidth,height=2.5 in]{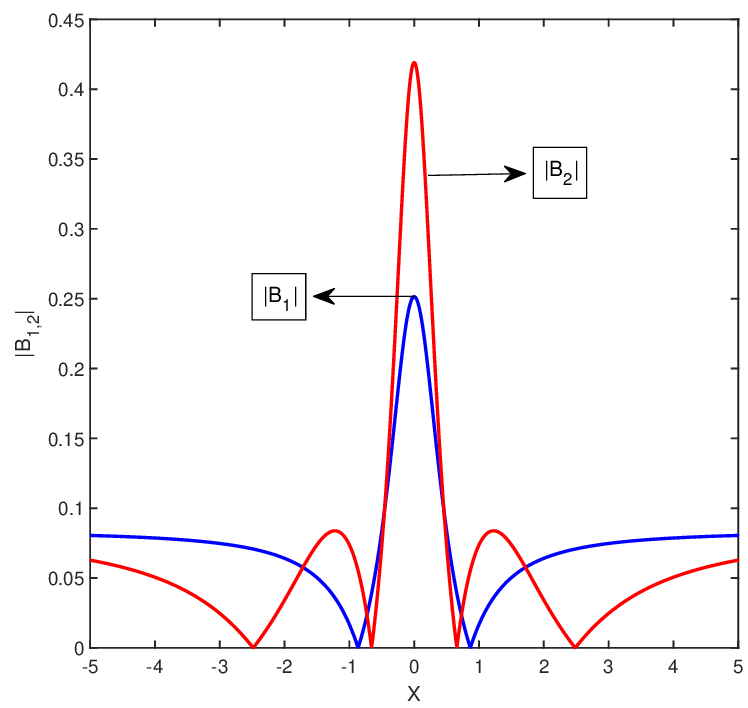}
     \caption{}
     \label{fig-1st-2nd-order-rogue-a}
 \end{subfigure}
 \hfill
 \begin{subfigure}[t]{0.49\textwidth}
     \centering
     \includegraphics[width=\textwidth,height=2.5 in]{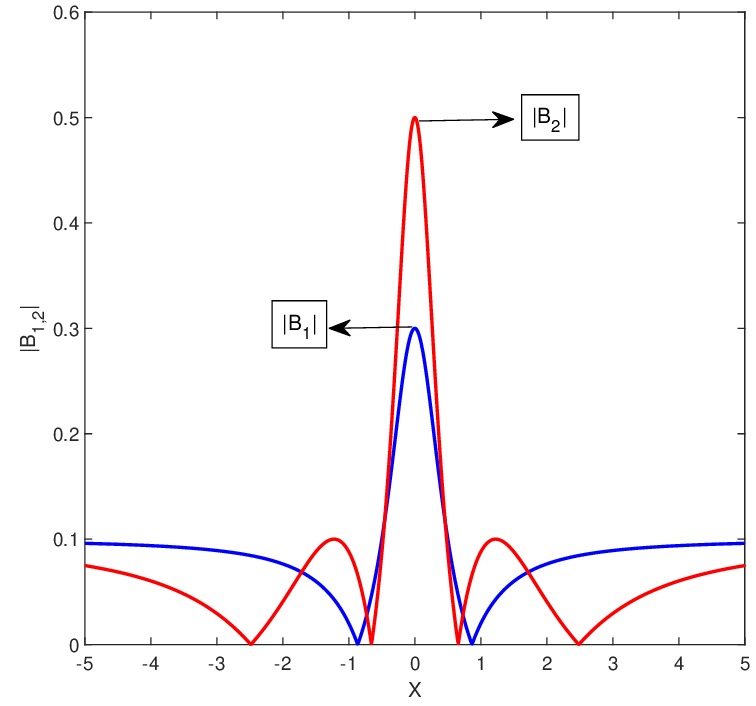}
     \caption{}
     \label{fig-1st-2nd-order-rogue-b}
 \end{subfigure}
	\caption{\justifying A comparison between first- and second-order rogue wave pulses $|B_1|,|B_2|$ is displayed [see Eqs. \eqref{eq-1st-order-rogue} and \eqref{eq-2nd-order-rogue}] against $X$ for different values of the rotational frequency $\Omega_0$ and angle of rotation $\theta$. The second-order correction of the rogue wave solution enhances the amplitude and width of the wave pulses. Compared to first-order rogue waves, second-order rogue waves are spikier because they absorb more energy from the surrounding waves. Here, $(C_g, C_c, \Omega_0, \theta)=(0.5, 0.2, 0.9, \pi/3)$ and  $(C_g, C_c, \Omega_0, \theta)=(0.1, 0.7, 0.9, \pi/3)$ in subplots (a) and (b), respectively. All other parameters remain unchanged  as in Fig. \ref{Fig-solitary-profile}.}
	\label{fig-1st-2nd-order-rogue}
\end{figure*}
%%%%%%%%%%%%%%%%%%%%%%%%%%%%%%%%%
To gain some insight, we analyse the effects of the parameters $C_g$ and $C_c$ related to thermal and cosmic ray pressures, respectively, as well as the impact of the Coriolis force due to the rotation of the cosmic fluids with the angular frequency ($\Omega_0$) and the angle of rotation ($\theta$), on both the first- and second-order rogue wave solutions \eqref{eq-1st-order-rogue} and \eqref{eq-2nd-order-rogue}. Initially, the typical first-order rogue wave profiles $|B_1|$ are plotted against $X$ and $T$ for two different values of $\Omega_0$ in Fig. \ref{fig-3d-rogue-wave}, keeping all other parameters fixed as mentioned in the figure caption. It has been observed that the amplitude and width of the rogue wave profile are reduced due to the enhancement of rotational frequency. The observation is quite natural, as the pulse cannot gain energy from the background wave. To provide more specific details on the effects of the essential parameters on the evaluation of rogue waves, we plot the first-order rogue wave profiles $|B_1|$  against $X$ at an instant $T =0$ for different values of $\Omega_0$, $\theta$ (subplots (a) in Fig. \ref{fig-1st-order-rogue}) and $C_g$, $C_c$ (subplots (b) of Fig. \ref{fig-1st-order-rogue}).  The figures illustrate that as the values of the parameters $\Omega_0$ and $\theta$ increase both the amplitude and width decrease, i.e., an increase in $\Omega_0$ and/or $\theta$ leads to the contraction of the rogue wave pulse. On the other hand, we notice that the inclusion of cosmic ray pressure with the thermal pressure significantly enlarges the rogue wave pulses. Subplot (b) in Fig. \ref{fig-1st-order-rogue} depicts that both the amplitude and width increase with the increasing value of $C_g$ and $C_c$ related to the thermal pressure and cosmic ray pressure, respectively. The results obtained indicate that, in the ISM of spiral galaxies, the temporal evaluation of rogue waves gains a significant amount of energy due to the modification of thermal pressure by cosmic ray pressure. Increasing rotational frequency leads to dissipation of
energy and decreases the rogue wave amplitude. The influence of cosmic rays amplified the rogue wave, which indicates the creation of regions of an enhanced magnetic field. These observations in Fig. \ref{fig-1st-order-rogue} are found to be similar for the second-order solution $|B_2|$ versus $X$ at $T=0$ as displayed in Fig. \ref{fig-2nd-order-rogue}. However, the second-order pulses' amplitude is higher than that of first-order rogue wave pulses.
The reduction (enhancement) of the amplitude and width of the rogue wave profiles, for both the first- and second-order solutions, signifies that the nonlinearity in the present model diminishes (increases) due to the increase in the values of $\Omega_0$ and/or $\theta$ ($C_g$ and/or $C_c$ ). Hence, the Coriolis force due to the rotation of the cosmic fluids, and the modified pressure due to the inclusion of cosmic ray pressure with thermal pressure exhibit stabilizing behaviour on first- and second-order rogue wave solutions. 

A comparison between the first- and second-order rogue wave solutions has been presented for different values of $\Omega_0$ and $\theta$ in Fig. \ref{fig-1st-2nd-order-rogue}. The figure illustrates that the second-order correction of the rogue wave solution leads to an enhancement of the amplitude and width of the wave pulses. It has been reported that the second-order rogue waves absorb more energy from the surrounding waves than the first-order rogue waves do, making them spikier than the first-order.

%%%%%%%%%%%%%%%%%%%%%%%%
\section{SENSITIVITY ANALYSIS} \label{sec-sensitivity}
In the present study, we have investigated the excitations of nonlinear magnetosonic solitons, shocks, and rouge waves in the ISM of spiral galaxies. The analytical and numerical studies of these nonlinear structures, presented in Secs.  \ref{sec-nls} and \ref{Sec-MI}, imply that the evaluation and characteristics of such waves largely depend on the parameters described in this model. We have observed from our obtained numerical results that a small change in some parameters may lead to a significant influence on the wave characteristics. In order to study the relative importance of different plasma parameters and identify the parameters that influence the magnetic field strength $B$ most, we perform a normalized forward sensitivity analysis \cite{dwivedi2022modeling,mandal2023dynamic}. To investigate sensitivity analysis, we compute the normalized forward sensitivity indices $S_{\mu}^{B}$ of $B$ for a parameter $\mu$ (where $\mu$ is either of the parameters $C_g, C_c, \Omega_0, \theta, \kappa_0, \eta_0$), as given below,
\begin{equation}
    S_{\mu}^{B}=\frac{\mu}{B} \times \frac{\partial B}{\partial \mu}.
\end{equation}
 We calculate $S_{\mu}^{B}$ for each plasma parameter at the typical normalized parametric value $(C_g, C_c, \Omega_0, \theta, \kappa_0, \eta_0)$ $=(0.7, 0.6, 0.8,\pi/7, 0.9, 0.02)$. The calculated values of the magnetic field strength sensitivity index corresponding to the soliton solution \eqref{eq-1soliton-solu}, the shock solution \eqref{eq-kdvb-sol}, and the first-order rogue wave solution \eqref{eq-1st-order-rogue} at $(\zeta, v_0)=(0.8,0.1)$, $\chi=50$, and $(X,T,k)=(0.4,0.2,0.4)$, respectively, are presented in Table II. When the sensitivity index $S_{\mu}^{B}$ is positive, it means that the magnetic field strength $B$ is increasing as a function of the related parameter $\mu$, when it is negative, it means that $B$ is decreasing. To illustrate, we obtain the sensitivity index $S_{C_c}^{B}=2.2748$ of $B$ for the parameter $C_c$ corresponding to the magnetosonic shocks, which represents an increase of $1\%$ in $C_c$ will increase the magnetic field $B$ by $2.2748\%$. However, $S_{C_g}^{B}=-2.0590$ indicates that $B$ decreases by $2.0590\%$ as $C_g$ increases by $1\%$. The values obtained in Table II reveal that, for the magnetosonic shock structure, $C_c$ appears to be the most sensitive parameter with a positive impact on the magnetic field strength $B$. The parameters $C_g$ and $\theta$ have a negative sensitive impact on $B$ corresponding to the shock solution, and all other parameters have a positive impact. 
The parameters $C_g$, $C_c$, $\Omega_0$, and $\theta$ have negative influences on the magnetic field strength of the magnetosonic soliton with $\theta$ as the most sensitive parameter. However, we find that these parameters have positive influences on the magnetic field strength of the magnetosonic rogue wave, of which $\Omega_0$ is the most sensitive parameter. 

\begin{table} 
\begin{tabular}{|c|ccc|}
  \cline{2-4}
  \multicolumn{1}{c|}{}&
 \multicolumn{1}{c}{Sensitivity} &
  \multicolumn{1}{c}{index ($S_{\mu}^{B}$)} & corresponding to 
  \\ \hline
  \makecell{Physical\\ Parameters ($\mu$)} & \makecell{Soliton\\ solution} & \makecell{Shocks\\ solution} & \makecell{Rogue\\ solution}\\
\hline
\makecell{Thermal\\ speed ($C_g$)} & $-0.2671$  & $-2.0590$ & $0.6774$ \\
\makecell{Cosmic ray\\ speed ($C_c$)} & $-0.1966$  & $2.2748$ &  $0.4977$ \\
\makecell{Rotational\\ frequency ($\Omega_0$)} & $-0.3912$ & $2.0109$ & $2.1591$\\ 
\makecell{Rotational\\ angle ($\theta$)} & $-1.9347$  & $-1.4392$ & $0.4576$\\ 
\makecell{Magnetic\\ resistivity ($\eta$)} & - & $0.1169$ & -  \\
\makecell{Cosmic ray\\ diffusivity ($\kappa$)} & - &$1.8379$ &- \\ \hline
\end{tabular}  
\caption{\justifying Calculated sensitivity indices ($S_{\mu}^{B}$) of magnetic field strength $B$ with respect to the physical parameters $\mu=C_g, C_c, \Omega_0, \theta, \kappa_0, \eta_0$ corresponding to the magnetosonic soliton, shock, and rogue wave solutions are displayed at the typical normalized parametric value $(C_g, C_c, \Omega_0, \theta, \kappa_0, \eta_0)$ $=(0.7, 0.6, 0.8,\pi/7, 0.9, 0.02)$.}
\end{table}
%%%%%%%%%%%%%%%%%%%%%%%%%%%%%%%%%%
%%%%
\section{CONCLUSIONS} \label{sec-result}
To conclude, we have investigated the evaluation of linear and nonlinear magnetosonic waves in the interstellar medium of spiral galaxies, which are the combination of thermal and cosmic fluid, taking into account the magneto-rotational effects of the fluid. The typical consistent parameters adopted in our numerical assessment are relevant in such fluid mediums of spiral galaxies \cite{gliddon1966gravitational,turi2022magnetohydrodynamic}. The cosmic fluid is found to have a reasonable contribution in modifying the thermal pressure in terms of the total pressure. Consequently, the dispersion properties, nonlinear wave evaluation, and modulation instabilities are modified. The effects of the parameters $C_g$ and $C_c$, related to the thermal and cosmic ray pressure, respectively, have been demonstrated. The Coriolis force due to the rotation of the plasma is found to have significant refinement in the wave propon frequency ($\Omega_0$) and rotation angle ($\theta$) vary. Further, cosmic ray diffusivity ($\kappa$) and magnetic resistivity ($\eta$) play a crucial role in evaluating shock waves. The significant results obtained are summarized as follows:

\begin{itemize}
    \item The linear analysis reveals that the dispersion relation (\ref{eq-linear-dis-rel}) remains the same as obtained in Turi and Misra\cite{turi2022magnetohydrodynamic} in the absence of both cosmic ray diffusivity ($\kappa$) and magnetic resistivity ($\eta$). The waves become unstable due to the increase in the values of the parameters $\eta$ and $\kappa$. However, interestingly, the damping caused by cosmic ray diffusivity and magnetic resistivity becomes negligible as one increases the rotating frequency. Wave damping is reduced by increasing the effects of the rotational effect in the plasma model. The damping is also modified due to the presence of cosmic fluid in the ISM of spiral galaxies by accelerating the damping rate $|\gamma|$. 
    \item To perform nonlinear analysis, we first derived a KdVB equation by employing the reductive perturbation technique. In solving the KdVB equation, several nonlinear wave shapes have been evaluated analytically, and numerically. The results indicate that the parameters of cosmic ray diffusivity ($\kappa_0$) and magnetic resistivity ($\eta_0$) are responsible for the formation of shock structures in the current model. When the effects of $\kappa_0$ and $\eta_0$ grow considerably, the dispersion effect becomes negligible compared to dissipation, and the monotonic shocks become stationary. Additionally, it has been noticed that the numerical solutions lead to oscillatory shock wave profiles, and growing values of $\kappa_0$, $\eta_0$ and rotation coefficients produce damped oscillatory waves with reduced amplitude. The KdVB equation becomes the KdV equation without the presence of $\eta_0$ and $\kappa_0$, consequently, the solitary wave evaluation has been reported. 
    
    \item To characterize the weakly nonlinear development of the envelope of a modulated wave packet in the low-frequency limit, we derive a nonlinear Schr\"{o}dinger (NLS) equation from Eq. \eqref{eq-kdvb}. It has been noticed that incorporating cosmic ray pressure in the present model reduces MI growth. As the modulated wave number $K$ increases, the growth rate rises until it hits a critical growth rate $\Gamma_c$, which varies depending on the plasma properties. It sharply declines for even higher values of $K$ after hitting the critical value growth rate. Furthermore, the present model supports the propagation of magnetosonic rogue waves. It is observed that the Coriolis force reduces the nonlinearity of the model, resulting in a shorter rogue wave amplitude, however, the inclusion of cosmic ray pressure with the thermal pressure significantly enlarges the rogue wave pulses. 
  
\end{itemize}
\par Our investigation is based on typical plasma parameters: $B_0 \sim 1$ nT, $\rho_0 \sim 10^{-21}$ Kg/m$^3$, $P_{g0}\sim 10^{-13}$ N/m$^2$, $P_{c0}\sim 10^{-13}$N/m$^2$, $\eta\sim 10^{-2} V_A^2/\omega_{ci}$ and $\kappa\sim 10^{-1} V_A^2/\omega_{ci}$ which are relevant to the spiral galaxies \cite{gliddon1966gravitational,turi2022magnetohydrodynamic} (See. Table I for more detail). The obtained results show that cosmic rays modify the existing pressure law and enhance the galactic magnetic field strength due to the interaction of cosmic rays with ionizied gas in the ISM clouds of spiral galaxies. The interplay between cosmic ray diffusivity and magnetic resistivity leads to the evaluation of shock waves in ISM of spiral galaxies. The higher values of these two parameters amplify the shock waves. Whereas, the lower values of these parameters lead to the formation of sloitons and rogue waves. The parameter related to the cosmic ray pressure in ISM clouds significantly modifies these wave structures. Thus, we believe the present study could be useful for understanding the magnetosonic wave propagation in the interstellar medium of spiral galaxies, where cosmic rays play an important role.
%%%%%%%%%%%%%%%%%%%%%%%%%%%%%%%%
\section*{Data availability statement}
Data sharing is not applicable to this article as no new data were created or analyzed in this study.

\section*{ACKNOWLEDGEMENTS}
One of us, J. Turi is grateful to the Council of Scientific and Industrial Research (CSIR) for a Senior Research Fellowship (SRF) with Ref. No. 09/202(0115)/2020-EMR-I. The authors thank Ms. Snehalata Nasipuri and Mr. Gourav Mandal for the valuable discussion to improve the work. The authors also thank the anonymous referee for his crucial and constructive suggestions to build up the work in its present form.

%By substituting into Eqs. ()–(6) and (10) and isolating distinct
%orders in $\epsilon$, we obtain the nth-order reduced equations:

%\bibliographystyle{plain}
\bibliographystyle{elsarticle-num}
\bibliography{ref1}
\end{document}